\documentclass[longauth]{aa} 

\usepackage{graphicx}
\usepackage{txfonts}
\usepackage[utf8]{inputenc}
\usepackage{array}
\usepackage{float}
\usepackage{multirow}
\usepackage[]{natbib}
\usepackage{amsmath}
\usepackage[colorlinks=true,linkcolor=blue,citecolor=blue]{hyperref}
\newcommand{\rev}[1]{#1} 

\begin{document} 

   \title{Testing giant planet formation in the transitional disk of SAO~206462 using deep VLT/SPHERE imaging\thanks{Based on observations collected at the European Organisation for Astronomical Research in the Southern Hemisphere under ESO programmes 095.C-0298 and 090.C-0443.}}

    \author{A.-L. Maire\inst{1}, T. Stolker\inst{2}, S. Messina\inst{3}, A. M\"uller\inst{1,4}, B.~A. Biller\inst{5,1}, T. Currie\inst{6}, C. Dominik\inst{2}, C.~A. Grady\inst{7}, A. Boccaletti\inst{8}, M. Bonnefoy\inst{9}, G. Chauvin\inst{9}, R. Galicher\inst{8}, M. Millward\inst{10}, A. Pohl\inst{1,11,12}, W. Brandner\inst{1}, T. Henning\inst{1}, A.-M. Lagrange\inst{9}, M. Langlois\inst{13,14}, M.~R. Meyer\inst{15,16}, S.~P. Quanz\inst{16}, A. Vigan\inst{14}, A. Zurlo\inst{17,18,14}, R. van Boekel\inst{1}, E. Buenzli\inst{16}, T. Buey\inst{8}, S. Desidera\inst{19}, M. Feldt\inst{1}, T. Fusco\inst{20}, C. Ginski\inst{21}, E. Giro\inst{19}, R. Gratton\inst{19}, N. Hubin\inst{22}, J. Lannier\inst{9}, D. Le Mignant\inst{14}, D. Mesa\inst{19}, S. Peretti\inst{23}, C. Perrot\inst{8}, J.~R. Ramos\inst{1}, G. Salter\inst{14}, M. Samland\inst{1,12}, E. Sissa\inst{19}, E. Stadler\inst{9}, C. Thalmann\inst{16}, S. Udry\inst{23}, and L. Weber\inst{23}
          }

   \institute{Max-Planck-Institut f\"ur Astronomie, K\"onigstuhl 17, D-69117 Heidelberg, Germany \\
          \email{maire@mpia.de}
         \and
         Anton Pannekoek Institute for Astronomy, University of Amsterdam, Science Park 904, 1098 XH Amsterdam, The Netherlands
         \and
	     INAF Catania Astrophysical Observatory, via S. Sofia 78, I-95123 Catania, Italy
         \and
         European Southern Observatory, Alonso de Cordova 3107, Casilla 19001 Vitacura, Santiago 19, Chile
         \and
         Institute for Astronomy, The University of Edinburgh, Royal Observatory, Blackford Hill View, Edinburgh, EH9 3HJ, UK
         \and
         National Astronomical Observatory of Japan, Subaru Telescope, Hilo, HI, USA
         \and
         Exoplanets and Stellar Astrophysics Laboratory, NASA Goddard Space Flight Center, Greenbelt, MD, USA
         \and          
		 LESIA, Observatoire de Paris, PSL Research University, CNRS, Universit\'e Paris Diderot, Sorbonne Paris Cit\'e, UPMC Paris 6, Sorbonne Universit\'e, 5 place J. Janssen, F-92195 Meudon, France
		 \and
         Univ. Grenoble Alpes, CNRS, IPAG, F-38000 Grenoble, France
         \and
         York Creek Observatory, Georgetown, Tasmania, Australia
         \and
         Heidelberg University, Institute of Theoretical Astrophysics, Albert-Ueberle-Str. 2, 69120 Heidelberg, Germany
         \and
         International Max Planck Research School for Astronomy and Cosmic Physics, Heidelberg, Germany
         \and
          CRAL, UMR 5574, CNRS/ENS-L/Universit\'e Lyon 1, 9 av. Ch. Andr\'e, F-69561 Saint-Genis-Laval, France
          \and
		  Aix Marseille Univ., CNRS, LAM, Laboratoire d'Astrophysique de Marseille, Marseille, France
		  \and
		  Department of Astronomy, University of Michigan, 1085 S. University  Ave, Ann Arbor, MI 48109-1107, USA
		  \and
          ETH Zurich, Institute for Astronomy, Wolfgang-Pauli-Strasse 27, 8093 Zurich, Switzerland
          \and
          N\'ucleo de Astronom\'ia, Facultad de Ingenier\'ia, Universidad Diego Portales, Av. Ejercito 441, Santiago, Chile
          \and
          Millennium Nucleus ``Protoplanetary Disk'', Departamento de Astronom\'ia, Universidad de Chile, Casilla 36-D, Santiago, Chile
          \and
          INAF--Osservatorio Astronomico di Padova, Vicolo dell'Osservatorio 5, I-35122 Padova, Italy
          \and
          ONERA, The French Aerospace Lab BP72, 29 avenue de la Division Leclerc, 92322 Ch\^atillon Cedex, France
          \and
		  Leiden Observatory, Leiden University, P.O. Box 9513, 2300 RA Leiden, The Netherlands
		  \and
          European Southern Observatory, Karl Schwarzschild St, 2, 85748 Garching, Germany        
          \and
          Geneva Observatory, University of Geneva, Chemin des Maillettes 51, 1290 Versoix, Switzerland
             }

    \date{Received 14/10/2016; accepted 14/02/2017}
    
 
  \abstract
   {The SAO~206462 (HD~135344B) disk is one of the few known transitional disks showing asymmetric features in scattered light and thermal emission. Near-infrared scattered-light images revealed two bright outer spiral arms and an inner cavity depleted in dust. Giant protoplanets have been proposed \rev{to account for} the disk morphology.}
   {We aim to search for giant planets responsible for the disk features and, in \rev{the} case of non-detection, to constrain recent planet predictions using the data detection limits.}
   {We obtained new high-contrast and high-resolution {total intensity images of the target spanning the $Y$ to the $K$ bands (0.95--2.3~$\muup$m) using the VLT/SPHERE near-infrared camera and integral field spectrometer}.}
  {The spiral arms and the outer cavity edge are revealed at high resolutions and sensitivities without the need for aggressive image post-processing techniques, which introduce photometric biases. We do not detect any close-in companions. For the derivation of the detection limits on putative giant planets embedded in the disk, we show that the knowledge of the disk aspect ratio and viscosity is critical for the estimation of the attenuation of a planet signal by the protoplanetary dust because of the gaps that these putative planets may open. Given assumptions on these parameters, the mass limits can vary from $\sim$2--5 to $\sim$4--7 Jupiter masses at separations beyond the disk spiral arms. The SPHERE detection limits are more stringent than those derived from archival NaCo/$L^{\prime}$ data and provide new constraints on a few recent predictions of massive planets (4--15~$M_{\rm{J}}$) based on the spiral density wave theory. The SPHERE and ALMA data do not favor the hypotheses on massive giant planets in the outer disk (beyond 0.6$''$). There could still be low-mass planets in the outer disk and/or planets inside the cavity.}
   {}


\authorrunning{A.-L. Maire et al.}
\titlerunning{Testing giant planet formation in the transitional disk of SAO~206462 using deep VLT/SPHERE imaging}

   \maketitle
%
\section{Introduction}

SAO\,206462 (HD\,135344B) is a rapidly rotating ($v$ sin $i$ $\sim$ 83 km\,s$^{-1}$) Herbig Ae/Be F4Ve star of age 9$\pm$2~Myr and mass 1.7$^{+0.2}_{-0.1}$~$M_{\odot}$ \citep{Mueller2011} {located at 156$\pm$11~pc \citep{GaiaCollaboration2016}} in the Upper Centaurus Lupus star-forming region, known to harbor a transitional disk resolved in scattered light \citep{Grady2009, Muto2012, Garufi2013, Wahhaj2015, Stolker2016} \rev{and} in thermal emission, both at mid-IR \citep{Doucet2006, Maaskant2013} and (sub-)mm wavelengths \citep{Brown2009, Andrews2011, Lyo2011, Perez2014, Pinilla2015, vanderMarel2016a, vanderMarel2016b}. It is part of a {widely separated ($\sim$21$''$, i.e. $\sim$3300~au) binary} with SAO~206463 \citep{Coulson1995}. Transitional disks are associated with an intermediate stage of disk evolution, where the dust opacity has been reduced (perhaps tracing \rev{a} dip in the gas surface density) in the near- and mid-IR \citep[e.g.,][]{Strom1989, Espaillat2014}. {Several scenarios have been proposed \rev{that can explain} the gas and dust depletion \rev{such as} photoevaporation, dust grain growth, \rev{or ongoing} planet formation \citep{Hollenbach1994, Dullemond2005, Lin1993}}.

\begin{table*}[t]
\caption{Observing log.}
\label{tab:obs}
\begin{center}
\begin{tabular}{l c c c c c c c c}
\hline\hline
UT date & Seeing ($''$) & $\tau_0$ (ms) & AM start/end & Mode & Bands & DIT\,(s)\,$\times$\,Nfr & $\Delta$PA ($^{\circ}$) & SR \\
\hline
2015/05/15 & 0.45--0.57 & 7--10 & 1.04--1.03 & IRDIFS\_EXT & $YJH$+$K1K2$ & 64$\times$64 & 63.6 & 0.78--0.85 \\ 
\hline
\end{tabular}
\end{center}
\tablefoot{The columns provide the observing date, the seeing and coherence time measured by the differential image motion monitor (DIMM) at 0.5~$\muup$m, the airmass at the beginning and the end of the sequence, the observing mode, the spectral bands, the DIT (detector integration time) multiplied by the number of frames in the sequence, the field of view rotation, and the Strehl ratio measured by the AO system.}
\end{table*}

Measurements of the CO line profiles \citep{Dent2005, Pontoppidan2008, Lyo2011} and of the stellar rotation \citep{Mueller2011} for SAO~206462 are consistent with an almost face-on geometry for the system ($i$\,$\sim$\,11$^{\circ}$). The star is actively accreting \citep[$\sim$10$^{-11}$--10$^{-8}$~$M_{\odot}$/yr,][]{GarciaLopez2006, Grady2009}. The disk is composed of a {massive outer component \citep[up to $\sim$2$''$ i.e. $\sim$300~au,][]{Lyo2011} with masses for the dust $M_{\mathrm{dust}}$\,$=$\,1.3$\times$10$^{-4}$~$M_{\odot}$ and gas $M_{\mathrm{gas}}$\,$=$\,1.5$\times$10$^{-2}$~$M_{\odot}$ \citep{vanderMarel2016a}}, scattered-light spiral features {within $\sim$80~au} indicative of dynamical processes \citep{Muto2012}, a large inner sub-mm dust cavity \citep[{$\sim$50~au},][]{Andrews2011}, and a sub-au dust accretion disk \citep{Fedele2008, Carmona2014, Menu2015}. Recently, \citet{Stolker2016} presented SPHERE optical and near-IR polarimetric images showing shadowing of the outer disk by warping/perturbation of the inner disk and the (marginal) detection of scattered light {up to $\sim$160~au}.

{One plausible explanation} for the spiral features seen in the SAO~206462 system is that they might be driven by planets. Two approaches have been used \rev{to investigate} this hypothesis in the literature. The first approach consists in fitting analytical formulae derived for linear perturbations, i.e. one planet driving one spiral, to the observations. \citet{Muto2012} proposed two planets located beyond $\sim$50~au with masses of $\sim$0.5~$M_{\rm{J}}$. Recently, \citet{Stolker2016} used a similar approach assuming two planets located interior or exterior to the spirals and showed that planets in the latter configuration with separations $\gtrsim$100~au provided a better match to their shape. The second approach exploits hydrodynamical simulations, which can account for non-linear perturbations, i.e. one planet can drive more than one spiral \citep[e.g.,][]{Dong2015, Zhu2015, Juhasz2015, Pohl2015}. In particular, {some studies have predicted one massive ($\sim$4--15~$M_{\rm{J}}$) planet located exterior to the SAO~206462 spirals \citep{Fung2015,Bae2016,Dong2017}}.

\citet{Garufi2013} suggested that one planet of mass from 13 to 5~$M_{\rm{J}}$ located at \rev{17.5--20~au} could explain the increase \rev{in} the cavity size measured for the small dust grains (28$\pm$6~au) and for the large dust grains \citep[39 to 50~au,][]{Brown2009, Andrews2011, Lyo2011}\footnote{{All these studies assumed a system distance of 140~pc}.}. This planet could also explain the spiral features. They could not rule out several massive planets inside the cavity, although this hypothesis seems at odds with the presence of diffuse gas. They also rejected several mechanisms not involving perturbing planet(s) \citep[photoevaporation, dust grain growth, magnetorotational instability,][]{Hollenbach1994, Dullemond2005, Chiang2007} \rev{to explain} the disk cavity. They finally showed that the disk is globally highly stable \citep[see also][]{Stolker2016}. Nevertheless, they could not rule out gravitational interactions between SAO~206462 and SAO~206463 \rev{as exciting mechanism for the disk spirals,} if {the components} have highly eccentric orbits.

Recently, \citet{vanderMarel2016b} presented ALMA data at spatial resolutions of 0.16$''$, which show that the previously detected asymmetric millimeter dust ring \citep{Perez2014, Pinilla2015} consists \rev{of} an inner ring and an outer asymmetric feature. They also proposed a different hypothesis \rev{that can explain} the disk features detected in both scattered light and thermal emission, where a vortex creates one spiral arm and a planet inside the cavity produces the other spiral arm.

Studies tried to detect stellar and \rev{substellar} companions in the disk of SAO~206462. \citet{Pontoppidan2008} ruled out a stellar companion inside the cavity from the detection of CO gas. VLT/NaCo imaging at 1.75 and 2.12~$\muup$m {discarded} low-mass stellar companions within the gap region ($>$0.22 $M_{\odot}$ {beyond 14~au}) and \rev{brown dwarf} companions more massive than {$>$19~$M_{\rm{J}}$ beyond 70~au} \citep{Vicente2011}\footnote{{They assumed for the star a distance of 140~pc and an age of 8~Myr, as well as the BT-DUSTY evolutionary tracks of \citet{Allard2011}. Using the contrasts measured at the corresponding angular separations and assuming the evolutionary models, distance and age used in this paper, we derive mass limits of $>$0.34 $M_{\odot}$ beyond 16~au and $>$28~$M_{\rm{J}}$ beyond 80~au.}}. However, these limits are not sensitive to planetary-mass companions. 

We present in this paper new high-contrast images of SAO~206462 covering the spectral range 0.95--2.32~$\muup$m obtained with the instrument VLT/SPHERE \citep{Beuzit2008} as part of the SpHere INfrared survey for Exoplanets (SHINE; Langlois et al., in prep.). We describe the observations and the data reduction \rev{(Sects.~\ref{sec:obs} and \ref{sec:datareduc})}. Then, we discuss the detection of point sources (Sect.~\ref{sec:pointsources}), the observed disk features (Sect.~\ref{sec:disk}) and the detection limits on putative giant protoplanets with respect to predictions from the literature (Sect.~\ref{sec:detectionlimits}). Finally, we analyze the sensitivity of the detection limits to the protoplanetary dust opacity considering the presence of gaps opened by putative giant embedded planets (Sect.~\ref{sec:dustopacity}).

\section{Observations}
\label{sec:obs}

SPHERE \citep{Beuzit2008} is an extreme adaptive optics (AO) instrument dedicated to high-contrast and high-resolution imaging of young giant exoplanets and circumstellar disks. The AO system \citep{Fusco2006} includes a fast 41$\times$41-actuators wavefront control, pupil stabilization, differential tip tilt \rev{control,} and toric mirrors \citep{Hugot2012} for beam transportation to the coronagraphs \citep{Boccaletti2008c} and science instruments. {In this paper, we used the near-infrared science instruments, the infrared dual-band imager and spectrograph IRDIS \citep{Dohlen2008a} and the integral field spectrometer IFS \citep{Claudi2008, Antichi2009}. The focal plane masks of the near-infrared coronagraphs are located in the common path and infrastructure of SPHERE, whereas the Lyot stops are located in IFS and IRDIS and are optimized for each instrument.}

We observed SAO~206462 on 2015 May 15 UT in the IRDIFS\_EXT mode (Table~\ref{tab:obs}). In this mode, {IRDIS and IFS} are operated in parallel, with IRDIS observing in the $K12$ filter pair \citep[$\lambda_{K1}$\,=\,2.110~$\muup$m and $\lambda_{K2}$\,=\,2.251~$\muup$m, $R$\,$\sim$\,20,][]{Vigan2010} and IFS in the $YJH$ bands \citep[0.95--1.65~$\muup$m, $R$\,$\sim$\,33,][]{Claudi2008}. The observing conditions were good and stable (wind speed $\sim$6--10~m/s, see Table~\ref{tab:obs}). The star was imaged with an apodized pupil Lyot coronagraph \citep{Soummer2005} of diameter 185~mas \citep{Boccaletti2008c}. At the beginning and the end of the sequence, {we acquired two dedicated calibrations. First, we recorded} for flux calibration purposes unsaturated images of the star out of the coronagraphic mask and inserting a neutral density filter of average transmission $\sim$1/100\footnote{The transmission curves of the neutral density filters can be found in the User Manual available at \url{www.eso.org/sci/facilities/paranal/instruments/sphere/doc.html}.} in the optical path. {Then, we} obtained coronagraphic images with four crosswise faint replicas of the star artificially generated using the deformable mirror \citep{Langlois2013} to measure the star location {for frame registering (Sect.~\ref{sec:datareduc}).} After the observation, we measured sky backgrounds. However, because of a hardware issue, the second unsaturated PSF image and the sky background images for IRDIS were not recorded properly. As a consequence, we used for the sky subtraction the closest \rev{sky backgrounds (taken the previous night)} obtained with the same filter and coronagraph configuration \rev{as} our observations and scaled them to the DIT and the background measured at large separation in the science images. {This non-optimal sky subtraction can affect the accuracy of the background subtraction to values below $\sim$8\% for the IRDIS RDI images, which are used \rev{to measure} the disk photometry} (Sect.~\ref{sec:disk}). All the other calibration data (darks, detector flats, wavelength calibration, IFU flats) were obtained during the following day. 

\begin{figure}[t]
\centering
\includegraphics[trim = 0mm 0mm 0mm 0mm,clip,width=0.32\textheight]{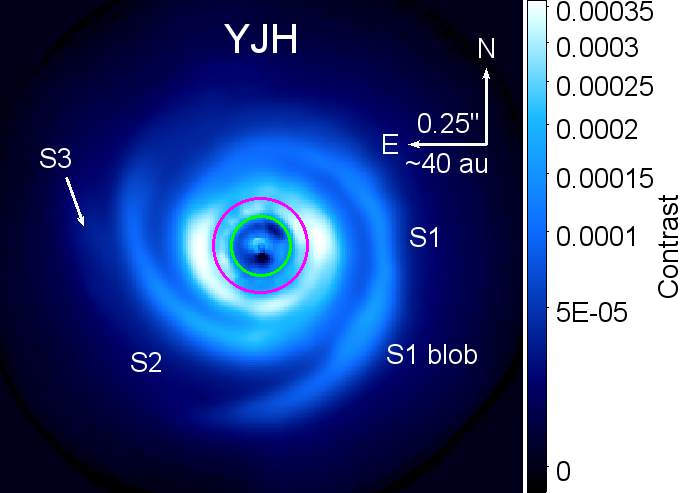}
\includegraphics[trim = 0mm 0mm 0mm 0mm,clip,width=0.32\textheight]{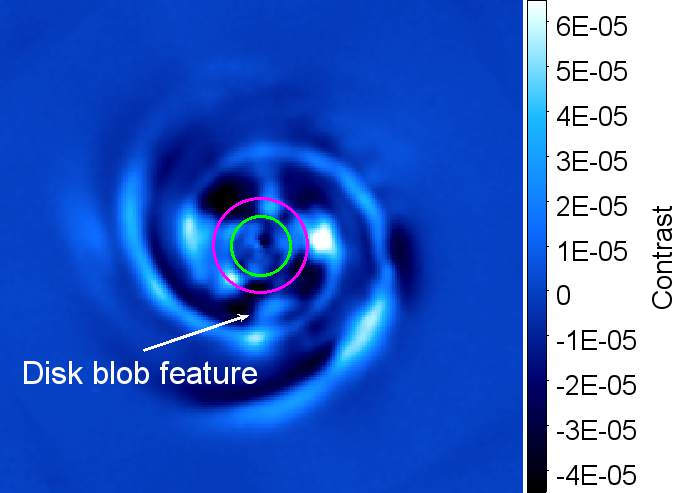}
\caption{{SPHERE/IFS images in the $YJH$ bands normalized to the {unsaturated non-coronagraphic PSF peak} after RDI (\textit{top panel}) and classical ADI (\textit{bottom panel}). The intensity scale is square root for the RDI image and linear for the ADI image. The green and magenta circles indicate the diameter of the coronagraph mask and regions dominated by the stellar residuals (diameter$\simeq$0.27$''$), respectively (see text).}}
\label{fig:ifsim}
\end{figure}

\section{Data reduction and analysis}
\label{sec:datareduc}

\begin{figure}[t]
\centering
\includegraphics[trim = 0mm 0mm 0mm 0mm,clip,width=0.31\textheight]{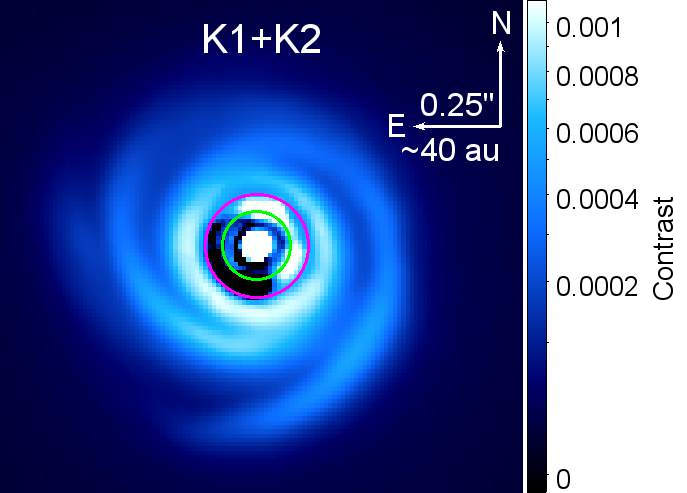}
\includegraphics[trim = 0mm 0mm 0mm 0mm,clip,width=0.31\textheight]{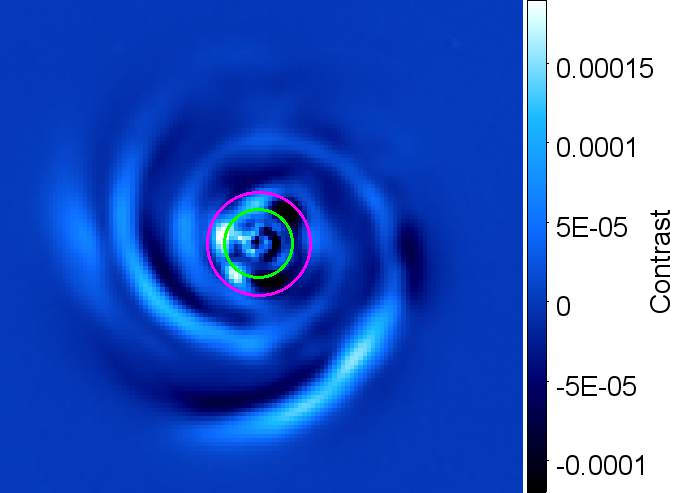}
\caption{{Same as Fig.~\ref{fig:ifsim} but for the SPHERE/IRDIS $K1+K2$-band images.}}
\label{fig:irdisim}
\end{figure}

\begin{figure*}[t]
\sidecaption
	\includegraphics[width=11.5cm]{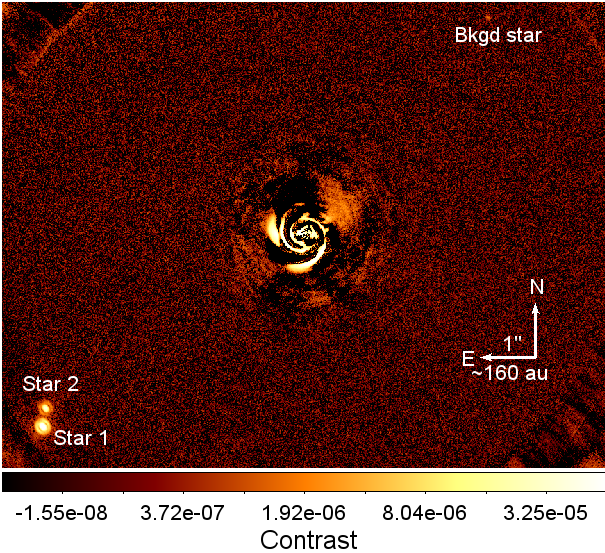}
     \caption{{Classical-ADI SPHERE/IRDIS $K1$-band image normalized to the maximum of the {unsaturated non-coronagraphic PSF} showing the close-in environment of SAO~206462 with the known stellar binary (noted Stars 1 and 2) and a well-separated point source classified as a background object ({labeled ``Bkgd star''}, see text). The intensity scale is logarithmic.}}
     \label{fig:irdisimfov}
\end{figure*}

The data were reduced with the SPHERE Data Center pipeline, which uses the Data Reduction and Handling software \citep[v0.15.0,][]{Pavlov2008} and dedicated IDL routines for the IFS data reduction \citep{Mesa2015}. The pipeline subtracts the sky background, corrects for the detector flat field, removes bad pixels, derives the IFS wavelength calibration, corrects for the IFU \rev{flat, corrects} for anamorphism \citep[0.60$\pm$0.02\%,][]{Maire2016a}, and registers the frames using the coronagraphic images taken with the satellite spots. We measured the full width at half maximum of the point-spread function (PSF) for the IRDIS images to be $\sim$56~mas ($K1$) and $\sim$59~mas ($K2$) and for the IFS images to be $\sim$37~mas ($J$) and $\sim$42~mas ($H$)\footnote{The temporal variations are below 0.2~mas. For IRDIS, we measured the variations on the individual frames recorded before the sequence. {The diffraction limits for SPHERE data are $\sim$31, $\sim$42, $\sim$54, and $\sim$58~mas in the $J$, $H$, $K1$, and $K2$ bands, respectively. Our data are thus diffraction-limited \rev{in} the $H$ and $K$ bands, but not in the $J$ band.}}. Then, the data were analyzed with a consortium image processing pipeline (R. Galicher, private comm.). This pipeline allows for several imaging post-processing techniques: classical angular differential imaging \citep[ADI,][]{Marois2006a}, subtraction of \rev{a radial profile from the data}, Locally Optimized Combination of Images \citep[LOCI,][]{Lafreniere2007a}, Template-LOCI \citep[TLOCI,][]{Marois2014}, and principal component analysis \citep[PCA, see][]{Soummer2012, Amara2012}. {We also considered reference differential imaging (RDI) in order to obtain unbiased views of the disk morphology. \rev{To build} the RDI reference, we {scaled in intensity the images of another star} without any disk features obtained in the same observing mode during the previous night. The quasi-static speckle pattern did not vary significantly between the two sequences, allowing for efficient rejection of the stellar residuals.} We show in Figs.~\ref{fig:ifsim} and \ref{fig:irdisim} the median-collapsed IFS and IRDIS images obtained with this method and with classical ADI and in Fig.~\ref{fig:irdisimfov} the full field of view of the IRDIS classical-ADI image. {Regions inside the magenta circles in Figs.~\ref{fig:ifsim} and \ref{fig:irdisim} were found to be dominated by the stellar residuals using the chromaticity of the stellar residuals.}

Three point sources are detected in the IRDIS field (Fig.~\ref{fig:irdisimfov}). Their photometry and astrometry were measured {using the TLOCI algorithm} applied to each spectral band separately. We divided each science frame into annuli of 1.5 full width at half maximum. Then, for each science frame and annulus, we computed a reference frame of the stellar residuals using the best linear combination of the 80 most correlated frames for which the {self-subtraction of mock point sources, \rev{modeled} using the observed PSF}, was at maximum 20\%. These parameters were selected from internal tests for the analysis of the consortium data. Negative synthetic companions \citep{Marois2010b, Bonnefoy2011} \rev{modeled} from the observed unsaturated PSF were inserted in the pre-processed data at the location of the detected point sources. We then processed the data assuming the TLOCI coefficients computed for the analysis without the synthetic companions. The \rev{subpixel} position and the flux of the modeled images were optimized to minimize the image residuals within a disk of radius 1.5 full width at half maximum (FWHM) centered on the measured companions \citep{Galicher2011a}. The error bars include the variations in the stellar flux during the sequence (estimated from the fluctuations of the stellar residuals, 0.009~mag for the $K1$ band) and the accuracy of the fitting procedure. The error term related to the variations in the PSF could not be estimated because no IRDIS PSF was recorded after the sequence (Sect.~\ref{sec:obs}). The astrometry of the three detected companions (Table~\ref{tab:photometryastrometry}) was calibrated using pixel scales of 12.267\,$\pm$\,0.009 and 12.263\,$\pm$\,0.009~mas/pix for the $K1$ and $K2$ filters respectively, and a \rev{north} angle offset of $-1.712$\,$\pm$\,0.063$^{\circ}$\footnote{We also accounted for offsets of $-$135.99\,$\pm$\,0.11$^{\circ}$ for the derotator zeropoint and $+$100.48\,$\pm$\,0.10$^{\circ}$ for the IFS field orientation with respect to the IRDIS field {\citep{Maire2016b}}.} \citep{Maire2016b}.

We finally used TLOCI \rev{to derive} the S/N maps for point sources. Each pixel value was divided by the standard deviation of the flux measured in \rev{an annulus of 1~FWHM} at the same angular separation. The TLOCI throughput was assessed using synthetic companions inserted at regular separations between 0.15$''$ and 6$''$. We iterated the TLOCI analysis on several position angles to average the effects of random speckle residuals. The azimuthally averaged contrast curves discussed in Sect.~\ref{sec:detectionlimits} were estimated from these S/N maps. {We note that although the disk signal is strongly attenuated by the TLOCI algorithm (which was optimized for point-source detection) in the SPHERE and NaCo data (the latter are described in Sect.~\ref{sec:nacodata}), the disk residuals increase somewhat the measured noise level. This effect is stronger in the SPHERE data, because of the higher S/N detection of the disk. However, these effects are intrinsic to the data and do not have to be corrected for.} {Finally, we corrected the SPHERE detection limits for the coronagraph transmission radial profile (A. Boccaletti, priv. comm.) and the small sample statistics following the prescription in \citet{Mawet2014}.}

\begin{table}[t]
\caption{{Astrometry and photometry relative to the star of the point sources detected in the IRDIS field of view (Fig~\ref{fig:irdisimfov}).}}
\label{tab:photometryastrometry}
\begin{center}
\begin{tabular}{l c c c c}
\hline\hline
 & Filter & $\rho$ (mas) & $\theta$ (deg) & $\Delta$mag \\
\hline
Star 1 & $K1$ & 5723$\pm$6 & 126.13$\pm$0.14 & 9.13$\pm$0.04 \\
 & $K2$ & 5723$\pm$6 & 126.14$\pm$0.14 & 9.26$\pm$0.03 \\
Star 2 & $K1$ & 5499$\pm$6 & 123.70$\pm$0.14 & 10.91$\pm$0.03 \\
 & $K2$ & 5499$\pm$6 & 123.70$\pm$0.14 & 10.98$\pm$0.03 \\
{Bkgd} & $K1$ & 5035$\pm$10 & 319.84$\pm$0.17 & 14.55$\pm$0.10 \\
{star} & $K2$ & 5038$\pm$19 & 319.88$\pm$0.25 & 14.55$\pm$0.23 \\
\hline
\end{tabular}
\end{center}
\tablefoot{The astrometric error bars were derived assuming an error budget including the error fitting, the uncertainties in the estimation of the location of the star \citep[1.2~mas from observations of bright stars,][]{Vigan2016}, pixel scale, north angle offset, and instrument distortion, and the accuracy of the dithering procedure \citep[0.74~mas,][]{Vigan2016}. The photometric error bars do not include the temporal variations of the PSF because only one PSF was relevant for the data analysis (see Sect.~\ref{sec:obs}).}
\end{table}

We note in the IFS ADI image (bottom panel of Fig.~\ref{fig:ifsim}) a blob feature interior to the S2 spiral south of the star at $\sim$0.20$''$ at the $\sim$3~$\sigma$ level\rev{, excluding the pixels at the same separation dominated by the disk features}. The feature is visible at the same location in all the individual IFS images and also in the images obtained after subtraction of a radial profile\rev{, but} not in aggressive PCA and TLOCI reduced data (results not shown). We conclude that this feature is likely a residual from a disk feature.

\section{Detection of point sources}
\label{sec:pointsources} 

\begin{figure}[t]
\centering
\includegraphics[trim = 8mm 0mm 2mm 0mm, clip, width=0.42\textwidth]{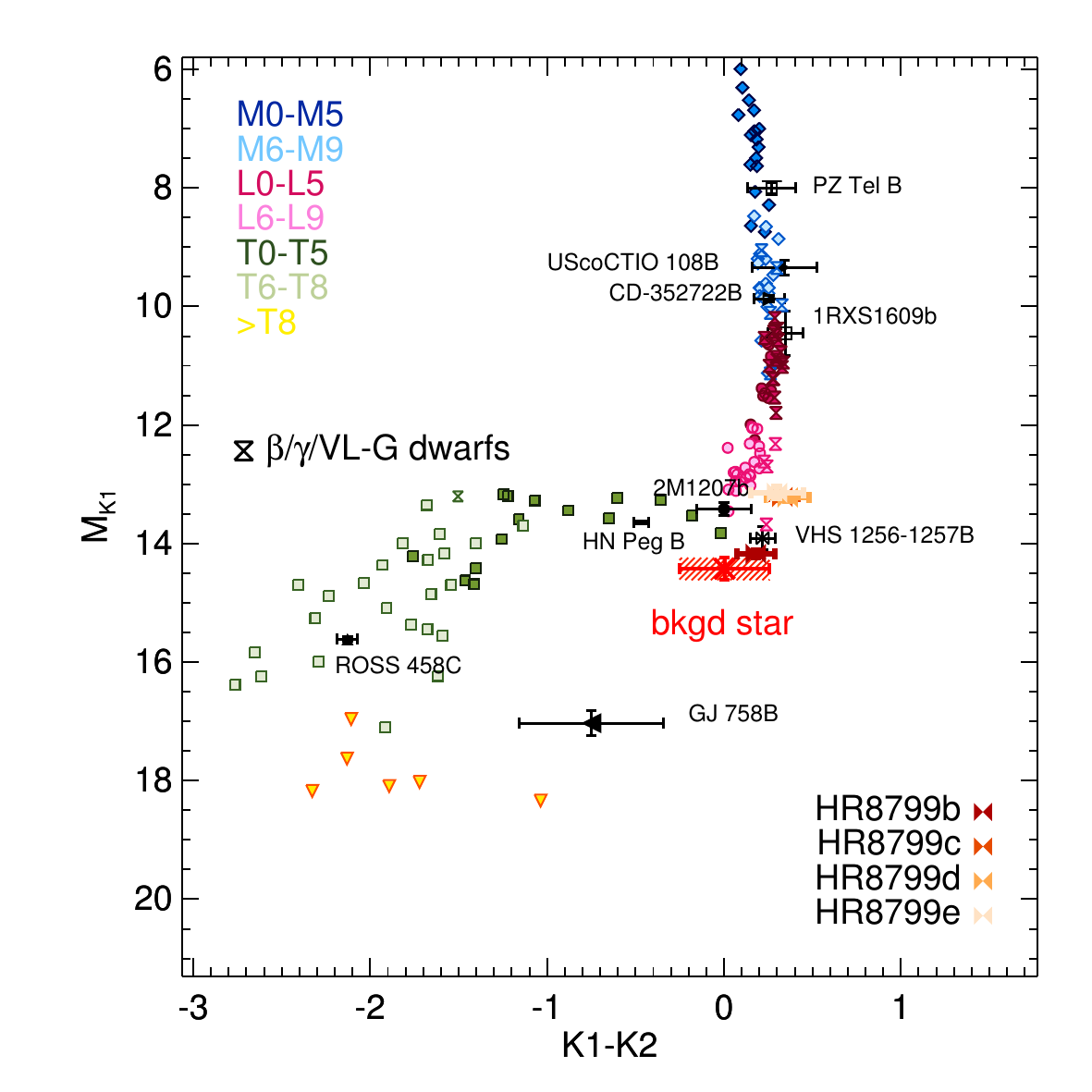}
\caption{{Color-magnitude diagram of the {point source noted ``bkgd star'' in Fig.~\ref{fig:irdisimfov} and Table~\ref{tab:photometryastrometry}}. Field dwarfs of M, L, and T types, very low-surface gravity dwarfs, and a sample of directly-imaged brown-dwarf and giant planet companions are also shown for comparison. {Note that the point source is subsequently classified as a background object from a common proper motion test (see Sect.~\ref{sec:pointsources}), hence its distance is unlikely to be the same as for SAO~206462.}}}
\label{fig:cmdiagram}
\end{figure}

\begin{figure}[t]
\centering
\includegraphics[trim = 6mm 8mm 12mm 12mm, clip, width=0.42\textwidth]{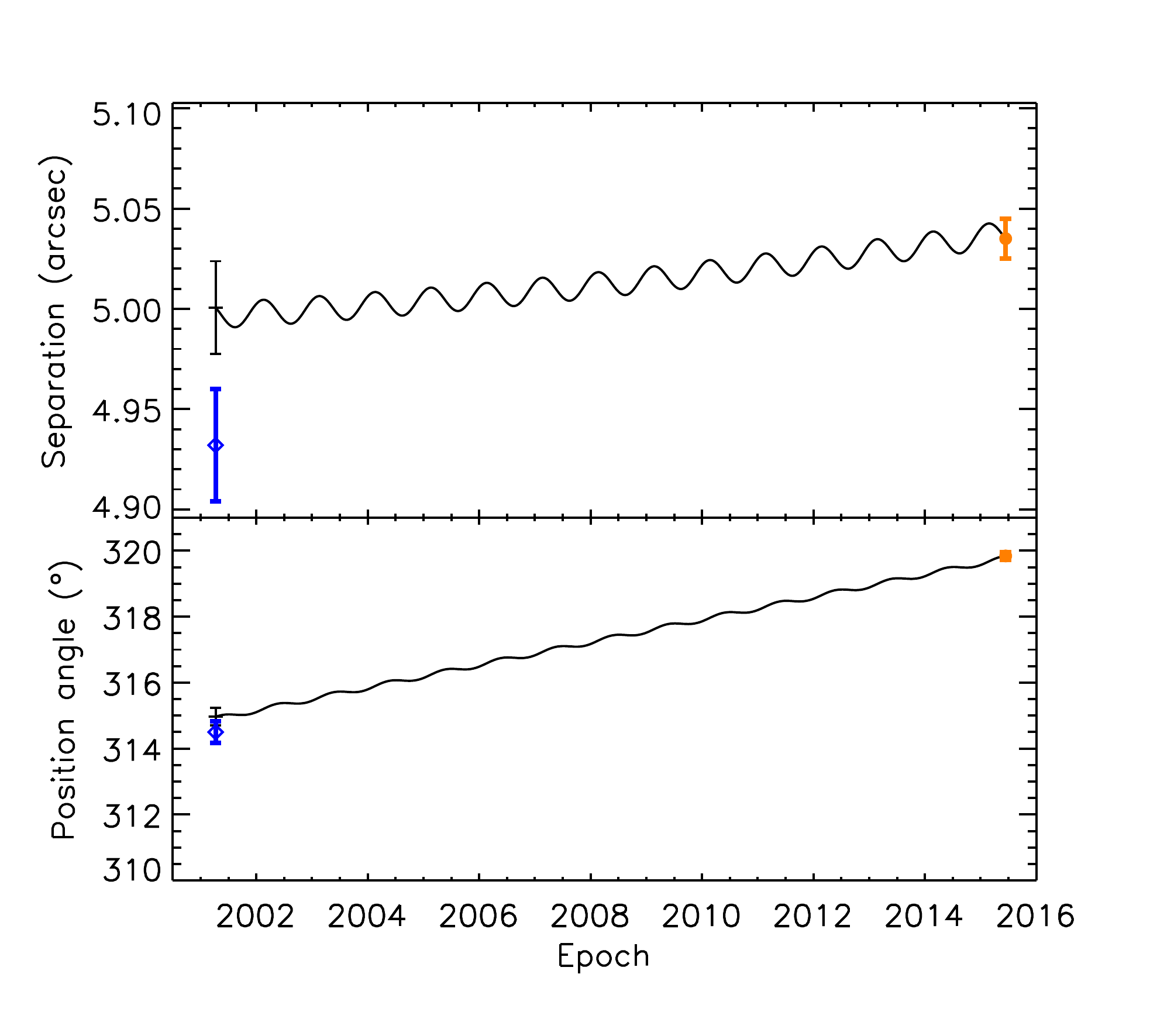}
\caption{{Relative astrometry of the {point source labeled ``bkgd star''} in Fig.~\ref{fig:irdisimfov} (colored data points with thick error bars) measured in SPHERE data (epoch 2015.37) and archival HST/STIS data (epoch 2001.27). The black cross represents the locations assuming that the point source is a stationary background object and accounting for the uncertainties in the proper motion and the distance of SAO~206462.}}
\label{fig:cpmtest}
\end{figure}

The two first point sources listed in Table~\ref{tab:photometryastrometry} (\rev{Stars 1 and 2}) were classified as a background stellar binary by \citet{Grady2009}. For the last point source (\rev{denoted} ``bkgd star'' for background star), we derived a mass range of $\sim$2--4~$M_{\rm{J}}$ according to the {atmospheric and evolutionary models of \citet{Baraffe2015} and \citet{Baraffe2003}} considering the uncertainties on its measured magnitudes and the system distance and age \citep{Mueller2011}. Using consortium tools developed for the classification and the ranking of the companion candidates discovered in SPHERE/SHINE, we also computed a background probability for this point source of $\sim$13.5\% assuming the Besan\c con model predictions \citep{Robin2003}. With the same suite of tools, we derived the color-magnitude diagram of the {point source} shown in Fig.~\ref{fig:cmdiagram} to compare its $K1-K2$ color to the colors of field and young dwarfs covering the M, L, T, and Y types. Further details about the derivation of the color-magnitude diagram are provided in {\citet{Mesa2016}. Its $K1-K2$ color} is similar to a late-L or early-T \rev{dwarf,} but its $K1$ magnitude is slightly fainter with respect to what is expected for objects of these spectral types. The point source is also seen in an HST/STIS image published in \citet{Grady2009} (program GO-8674, P.I. Lagrange). We retrieved \rev{the reduced image (corrected for cosmetics and distortion) from the HST archive,} and measured the astrometry of the {point source} ($r$\,=\,4932\,$\pm$\,28~mas, $\theta$\,=\,314.5\,$\pm$\,0.33$^{\circ}$) to compare it with the SPHERE astrometry. The results are shown in Fig.~\ref{fig:cpmtest}. The motion of the {point source} between the two epochs reflects the s{tellar proper motion \citep[$\mu_{\mathrm{\alpha}}$\,=\,$-$20.159\,$\pm$\,0.603~mas/yr, $\mu_{\mathrm{\delta}}$\,=\,$-$22.481\,$\pm$\,0.719~mas/yr,][]{GaiaCollaboration2016}}\rev{; it} is consistent with an object not gravitationally bound to SAO~206462. Therefore, we conclude that the {point source} is a background star.

\section{Disk imaging and photometry}
\label{sec:disk}

We analyze \rev{the} morphology and photometry of the detected disk features using the spatial and spectral information provided by the SPHERE data. A full \rev{modeling} of the total intensity images of the disk will allow \rev{a more quantitative analysis of} the physical \rev{constraints,} but is beyond the scope of this paper.

\subsection{Disk features}

\begin{figure*}[t]
\centering
\includegraphics[trim = 8mm 0mm 5mm 0mm, clip, width=0.42\textwidth]{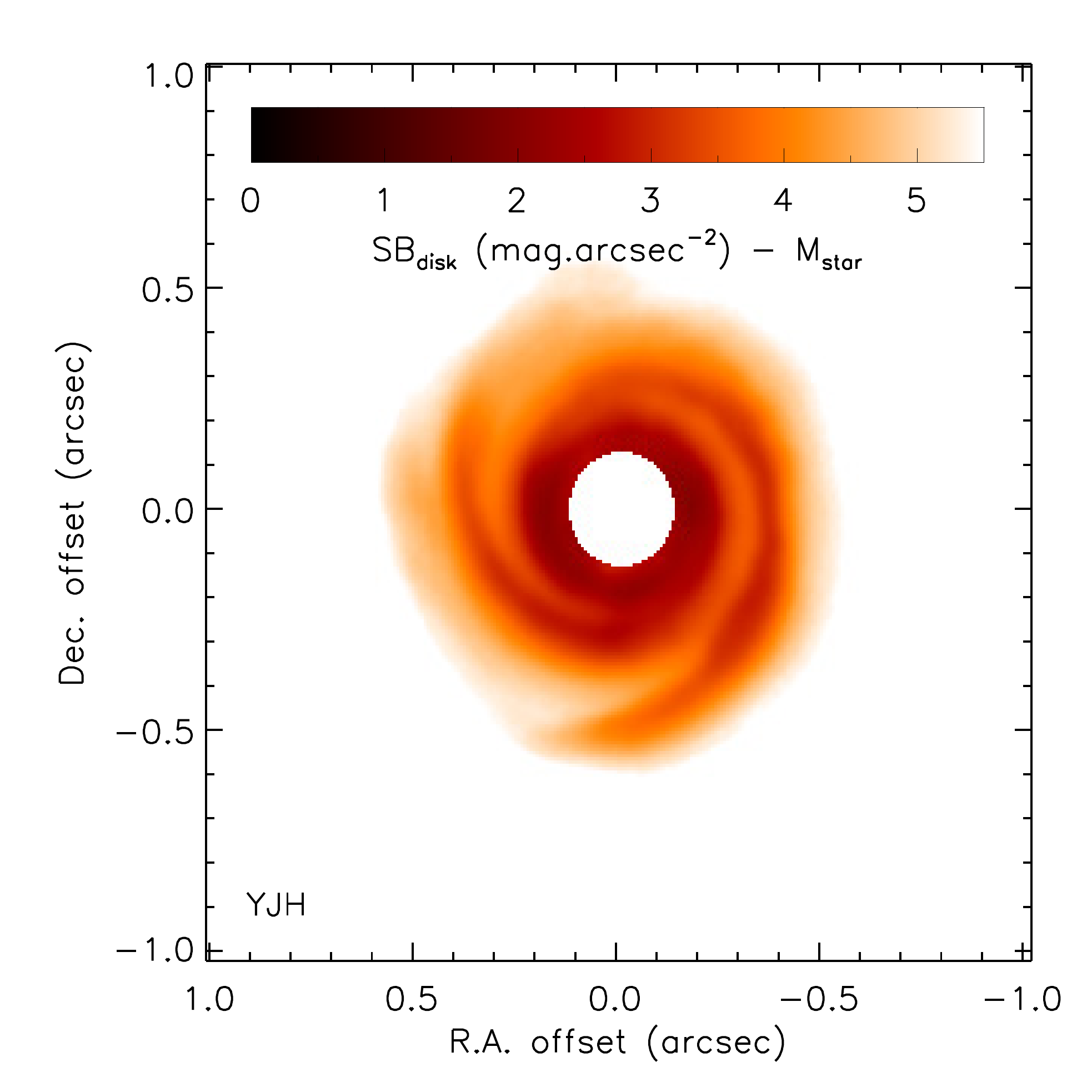}
\includegraphics[trim = 8mm 0mm 5mm 0mm, clip, width=0.42\textwidth]{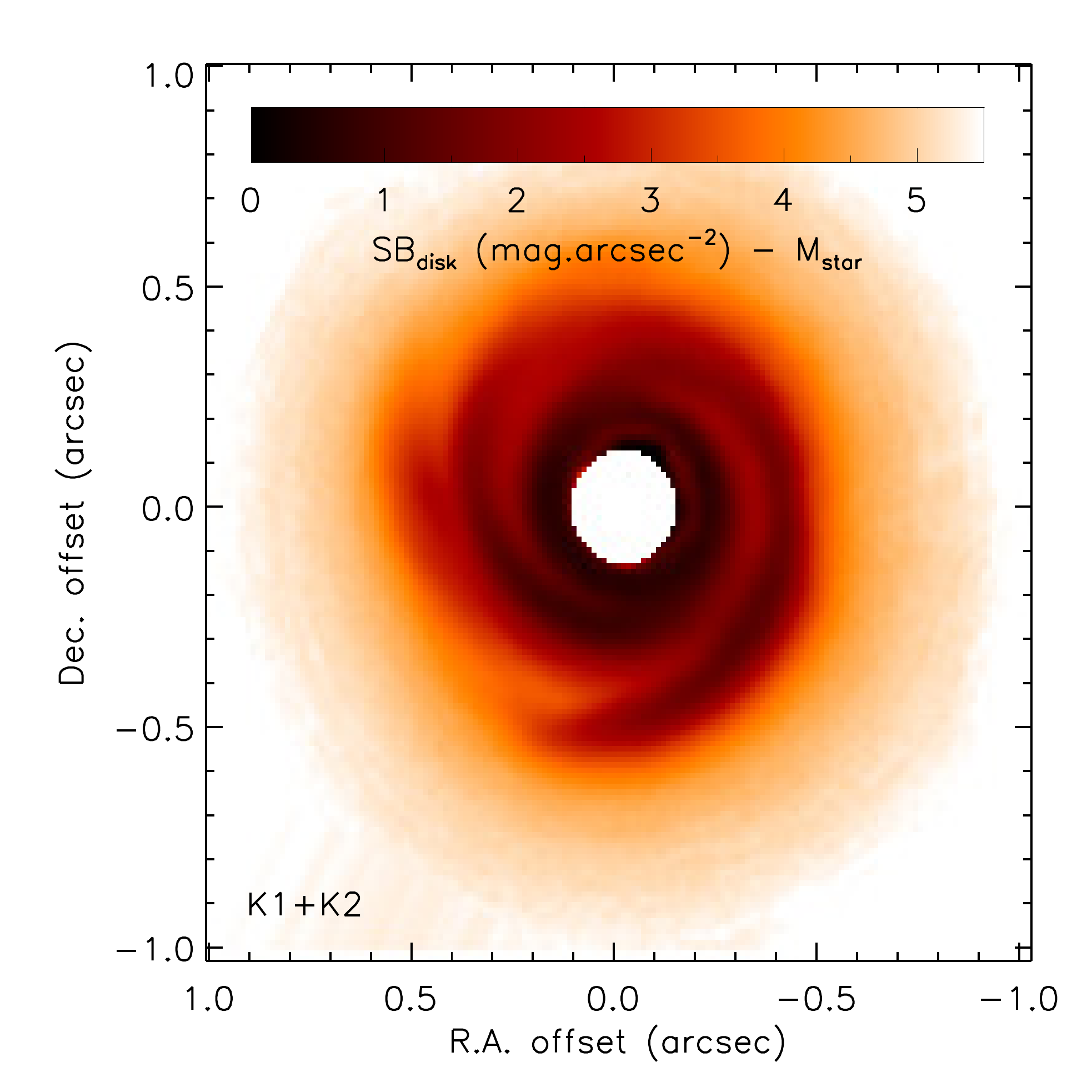}
\caption{{RDI IFS (\textit{left}) and IRDIS (\textit{right}) deprojected images collapsed over the spectral channels. The centers of the images have been masked out numerically in order to remove the star. North is up and East to the left.}}
\label{fig:spiralfit}
\end{figure*}

The spiral arms and the outer edge of the inner cavity of the disk are clearly detected in the SPHERE data {with non-aggressive differential imaging techniques} (top panels of Figs.~\ref{fig:ifsim} and \ref{fig:irdisim}, and Fig.~\ref{fig:spiralfit}), providing a nice illustration of the instrument capabilities for high-contrast imaging of faint extended sources such as circumstellar disks. {The disk is also \rev{clearly} seen in individual RDI IFS images} (Appendix~\ref{sec:ifsimchannels}). Except for an increase \rev{in} the disk bulk contrast with respect to the star, we do not see any clear evidence \rev{of} morphological spectral variations. The IRDIS and IFS images confirm that the disk spirals are launched axisymmetrically at the same corotation radius, as also found by \citet{Stolker2016}. We also detect in both IFS and IRDIS images the S3 spiral feature previously noted by \citet{Garufi2013} and \citet{Stolker2016} in the east part of the S2 spiral arm, especially at the longest wavelengths. We finally note that the outer rim looks brighter in the IRDIS $K$-band image (top panel of Fig.~\ref{fig:irdisim}).

We do not recover in the SPHERE images the radially extended spots to the northeast and the southwest seen in GPI total intensity $J$-band images by \citet{Wahhaj2015}. The latter suggested that they could be accretion streams from the disk into the star. Nevertheless, they mentioned that they could also be ADI and/or GPI artifacts because of the small field of view rotation of the data and \rev{the} presence of similar features in other GPI data \citep[e.g.,][]{Perrin2015}. Our dataset covers a field of view rotation 1.8 times larger \rev{than} the GPI data, which suggests that the features detected in the GPI data are probably artifacts.

\begin{figure}[!t]
\centering
\includegraphics[trim = 14mm 22mm 10mm 10mm, clip, width=0.42\textwidth]{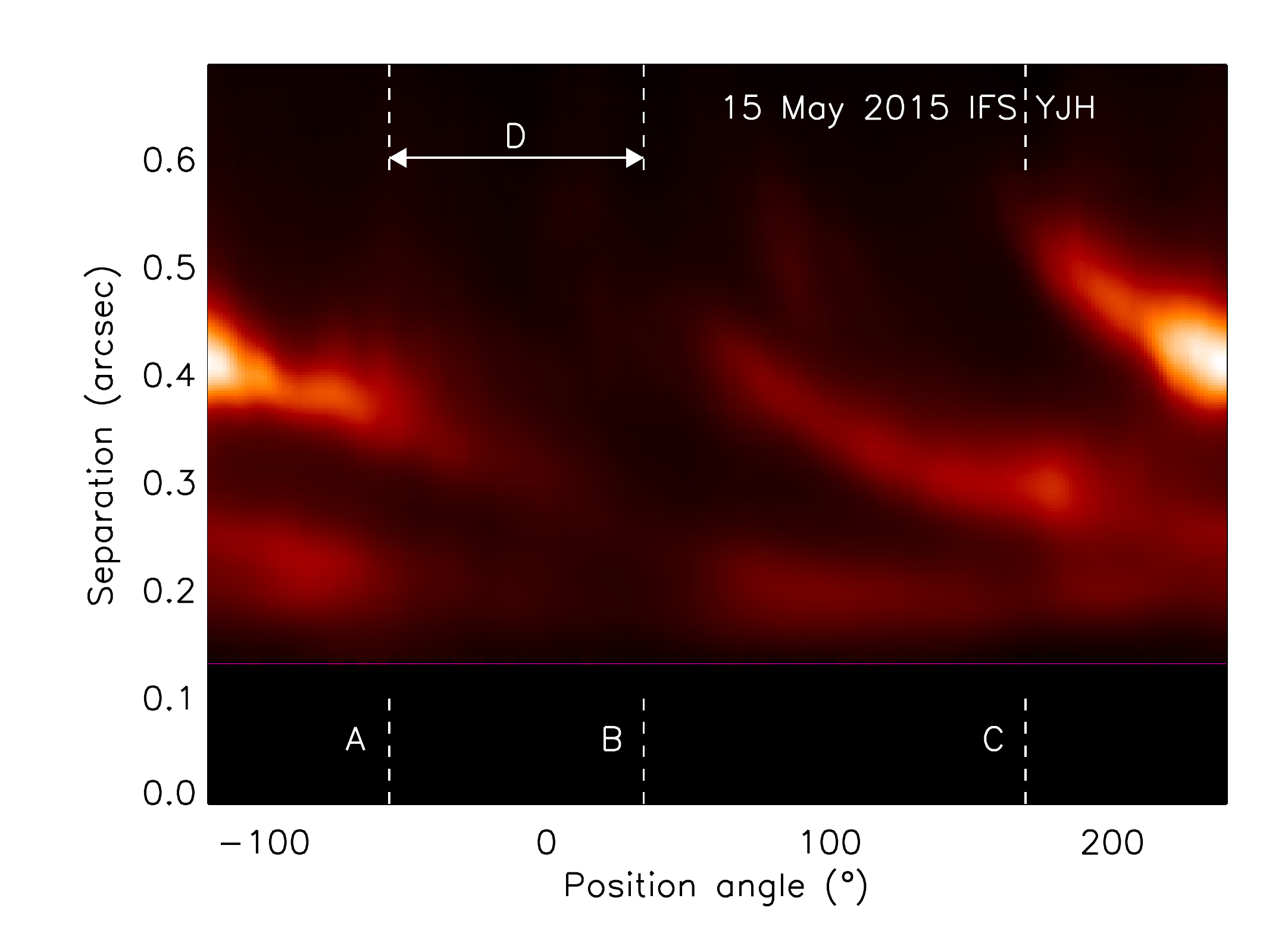}
\includegraphics[trim = 14mm 0mm 10mm 11mm, clip, width=0.42\textwidth]{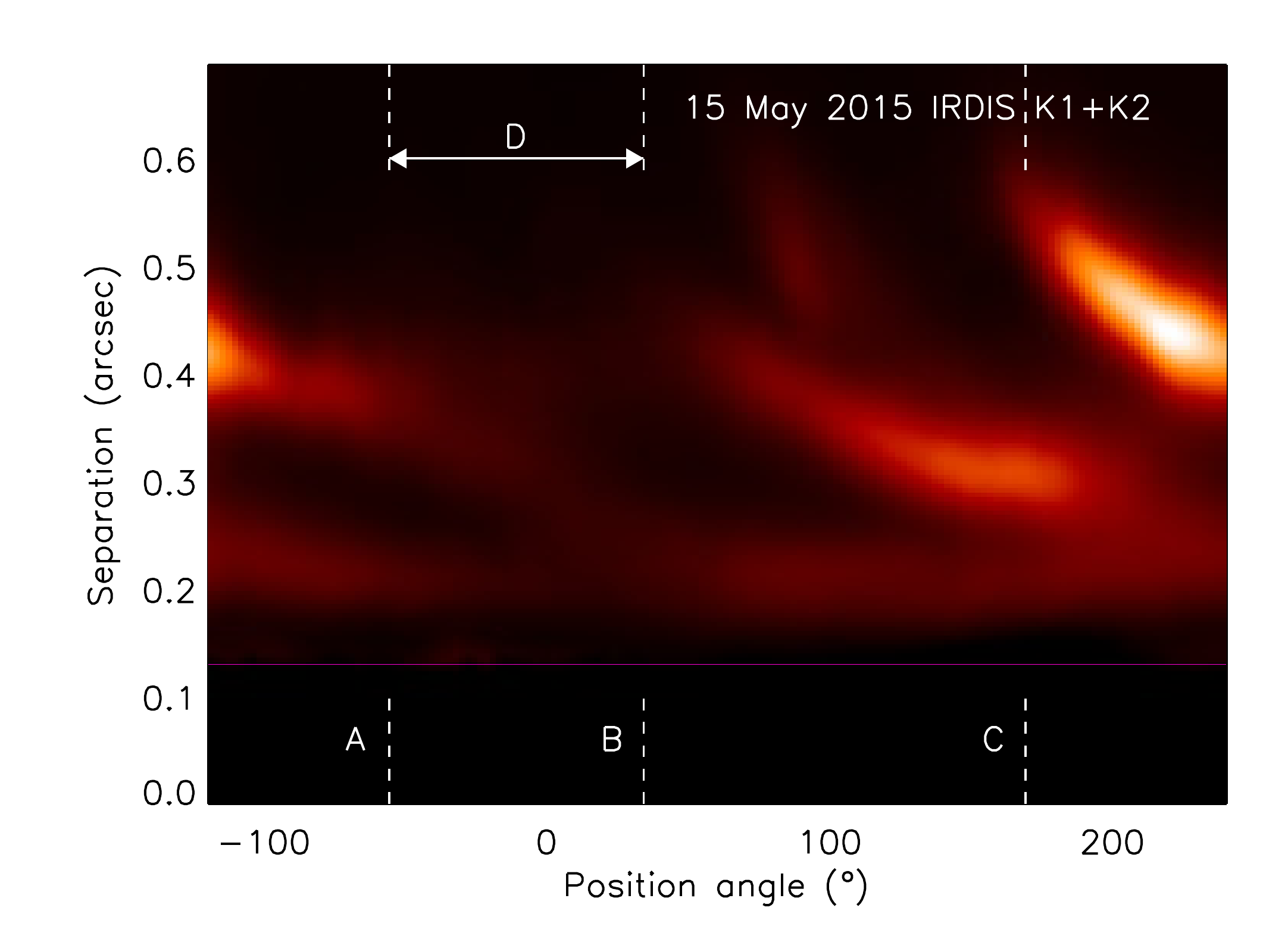}
\caption{{Polar projections of the deprojected RDI IFS (top) and IRDIS (bottom) images scaled by the distance to the star. We also show the locations of the shadow lanes found {and defined} by \citet{Stolker2016}. The intensity scale is different from the scale of Fig.~\ref{fig:spiralfit} in order to enhance the contrast of the shadow lanes. The centers of the images have been masked out numerically in order to remove the star (magenta lines).}}
\label{fig:polarproj}
\end{figure}

\citet{Stolker2016} detected shadow lanes in SPHERE polarized light images covering the $R$ to the $J$ bands (0.6--1.25~$\muup$m) obtained in late March and early May 2015. \rev{Figure}~\ref{fig:polarproj} shows the polar projections of the $r^2$-scaled RDI IFS and IRDIS images with the positions of the shadow lanes in the IRDIS $J$-band polarized image obtained on 2015 May 2, i.e. about two weeks prior to these observations. Features A, B, and D {\citep[notations from][]{Stolker2016}} are recovered in both polar-projected images, but feature C is only seen in the IFS image. Feature C was not seen in the SPHERE data obtained in late March 2015. The minima in the total intensity images are not as pronounced as in the polarized light images. This might be caused by sensitivity effects.

\subsection{Photometry and morphology of the spiral arms}

\begin{figure*}[t]
\centering
\includegraphics[trim = 12mm 0mm 4mm 0mm, clip, width=0.48\textwidth]{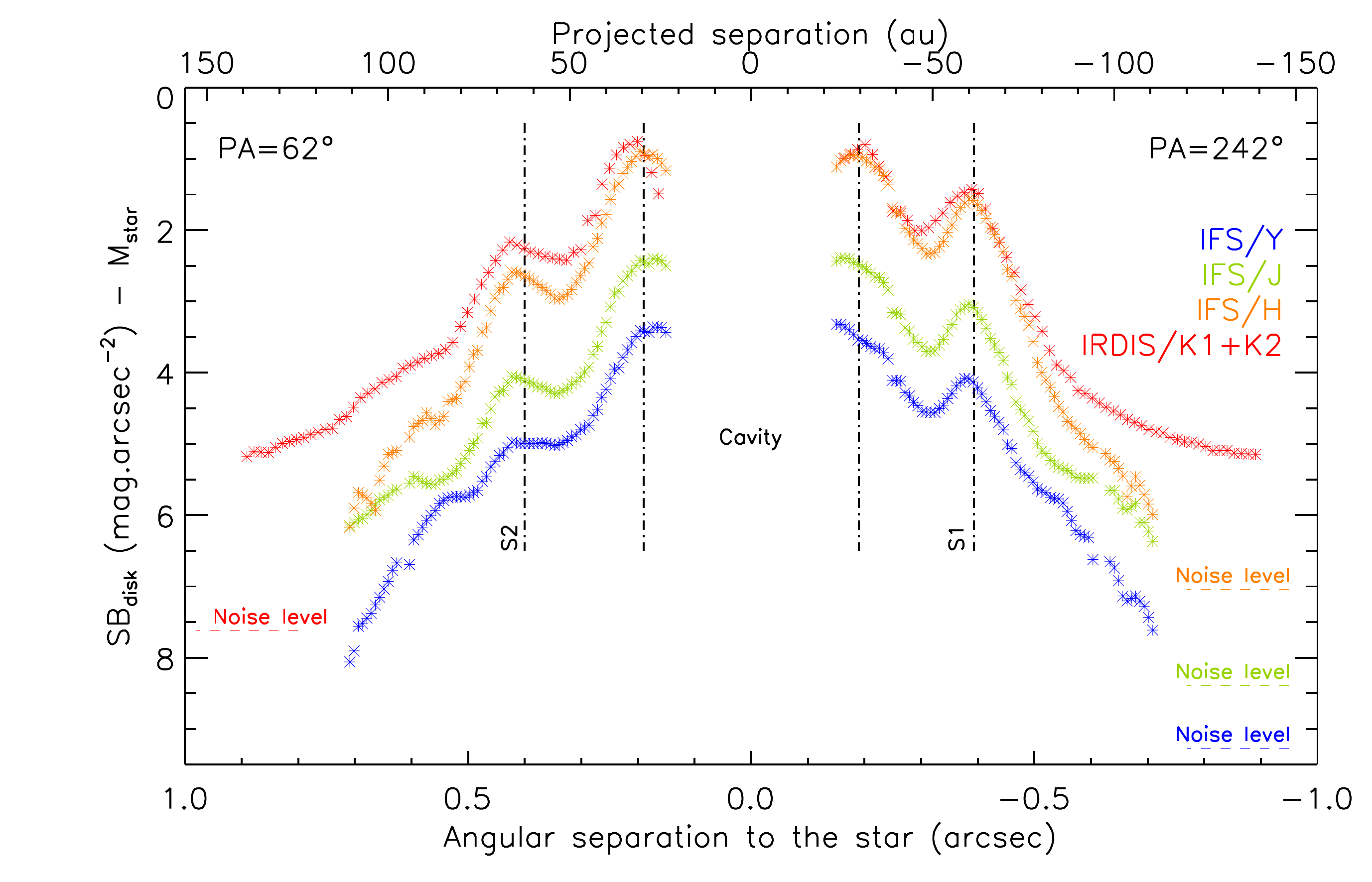}
\includegraphics[trim = 12mm 0mm 4mm 0mm, clip, width=0.48\textwidth]{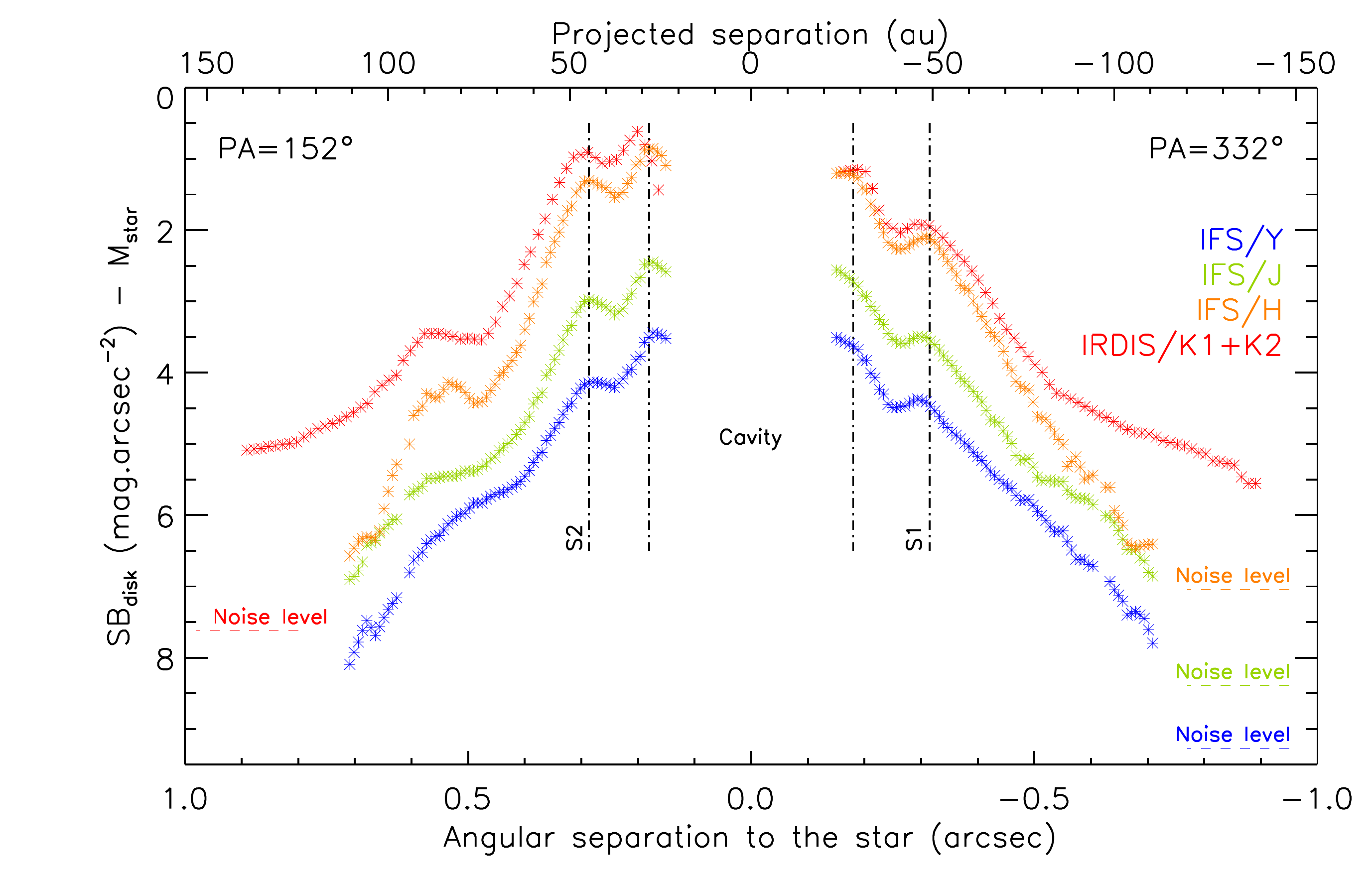}
\caption{{Radial surface brightness profiles along the major (\textit{left panel}) and minor (\textit{right panel}) axes of the disk for the RDI IFS and IRDIS images. {The profiles were corrected for the spectral dependency of the stellar flux (the label M$_{\mathrm{star}}$ on the $y$-axis refers to the stellar magnitude).} The noise levels at 1$\sigma$ are indicated as dashed horizontal lines (see text).}} 
\label{fig:brightnessprof}
\end{figure*}

We {used the RDI IFS and IRDIS images} (top panels of Figs.~\ref{fig:ifsim} and \ref{fig:irdisim}) to derive the radial surface brightness profiles of the disk along its major and minor axes (Fig.~\ref{fig:brightnessprof}), assuming that the major axis has a position angle of 62$^{\circ}$ \citep{Perez2014}. For IFS, we analyzed \rev{the $Y$, $J$, and $H$ spectral bands separately}. We used radial cuts through the disk of \rev{1 pixel in width}. The conversion of the intensity into mag/arcsec$^2$ was \rev{performed} using the 2MASS stellar magnitudes \citep[$J$\,=\,7.279, $H$\,=\,6.587, and $K$\,=\,5.843,][]{Cutri2003} and the ratio of the maximum to the total flux of the measured {unsaturated non-coronagraphic PSF}. With this normalization choice, {the profiles provide information on the scattering efficiency of the dust grains with the wavelength \citep[see][]{Quanz2012}}. We did not include the error on the photometric calibration. The noise levels (horizontal dashed lines) were estimated at large separation from the star.

{We note} \rev{the similar overall shapes of the profiles of} the spectral bands with bumps associated to the two spiral arms at the same locations. Since the spiral pattern is not centrosymmetric (the opening angle of S1 decreases abruptly by $\sim$15$^{\circ}$ at the location of {its bright blob feature}, cf. Figs.~\ref{fig:ifsim} and \ref{fig:irdisim}), the locations of the bumps are not centrosymmetric. The S/N of the disk detection drops quickly exterior to the spiral arms. The IFS radial profiles also show a sharp increase \rev{($\sim$2.5~mag)} from the $Y$ to the $H$ bands, which suggests an increasing efficiency of the scattering with the wavelength. An increasing trend (although less significant) was also seen in the polarized-light radial profiles from the $R$ to the $J$ bands presented in \citet{Stolker2016}. The IRDIS $K$-band profile does not seem to confirm the increasing trend observed for the IFS data. As discussed in \citet{Stolker2016}, the red colors of the dust grains could be explained if they are composed of $\muup$m-sized aggregates \citep{Min2016}.

\begin{figure}[t]
\centering
\includegraphics[trim = 10mm 0mm 6mm 7mm,clip,width=0.43\textwidth]{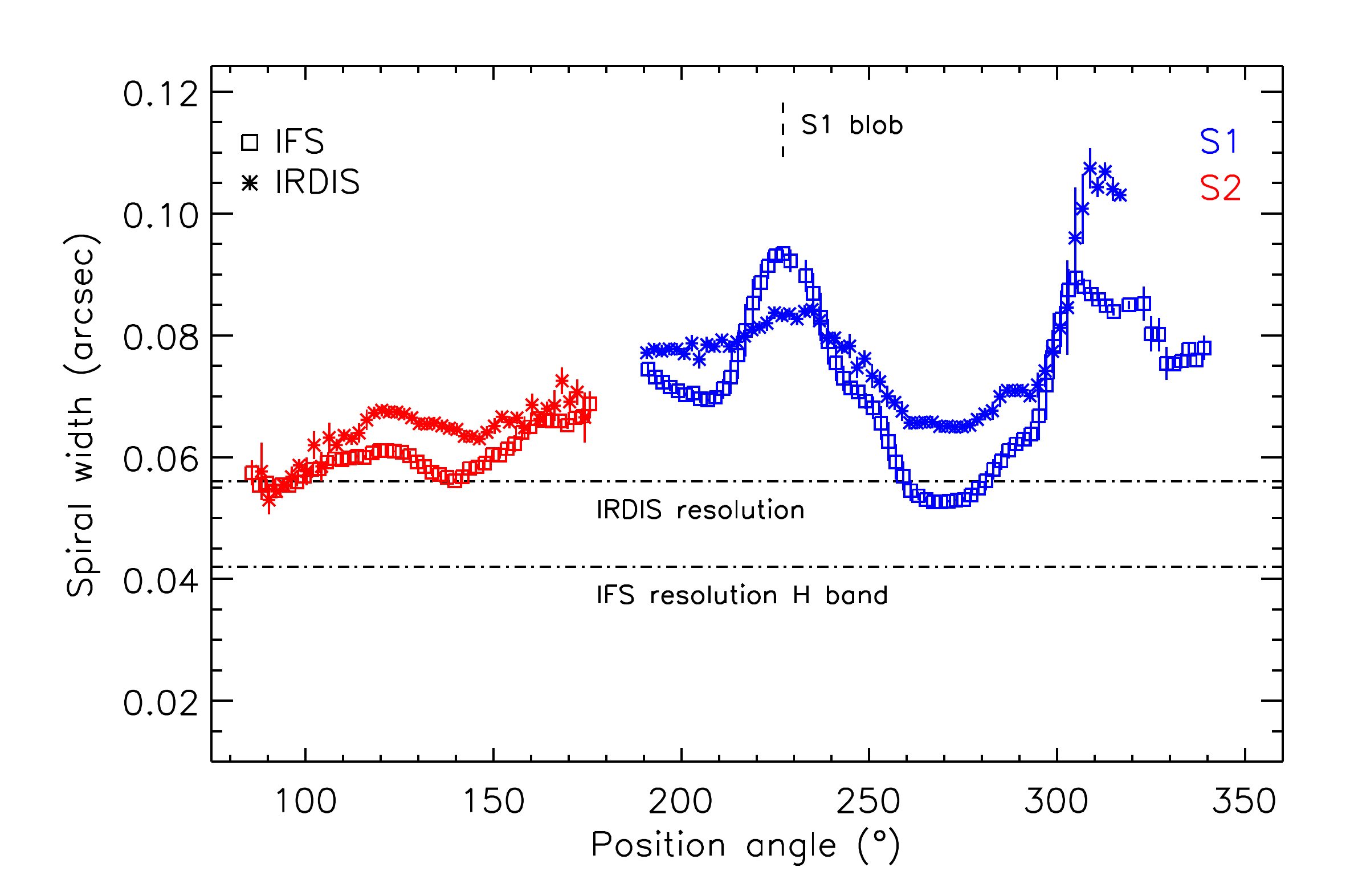}
\caption{{Widths of the spiral arms as a function of the {position angle} for both collapsed RDI IFS and IRDIS images (see text)}.}
\label{fig:widthspirals}
\end{figure}

Then, we registered the spine, photometry, and width of each spiral arm in the {collapsed RDI IFS and IRDIS images}. The images were first deprojected assuming for the disk an inclination of 11$^{\circ}$ \citep{Dent2005, Pontoppidan2008, Lyo2011} and a position angle of 62$^{\circ}$ \citep{Perez2014} and scaled by the square of the distance to the star. Then, \rev{we rotated the images by steps of 2$^{\circ}$ in} order to align the spiral parallel to the horizontal direction and \rev{to fit it with a 1D} Gaussian function using boxes of $\sim$0.15$''$ width (i.e., 20~pixels for IFS and 12~pixels for IRDIS) width approximately centered on the spiral spine to extract the location of the spine, the intensity at the spine location, and the width of the spiral. The error bars on the data points were assessed using their standard deviation in a \rev{three-bin} sliding box.

We \rev{show the width of the spirals as a function of the position angle in Fig.~\ref{fig:widthspirals}}. Each spiral arm exhibits \rev{a} similar overall shape \rev{in} the IFS and IRDIS images. As the position angle increases (i.e., going closer to the star), the IFS profile for S1 shows \rev{alternating} maxima and minima. {The peak-to-peak variations are $\sim$25--40~mas}. The IFS profile for S2 has a constant width of $\sim$60~mas in its external part then exhibits a slightly increasing slope for position angles larger than $\sim$140$^{\circ}$. We note some {local differences in the IRDIS profiles with respect to the IFS data that can be attributed to a poorer spatial resolution or sensitivity effects} (shallower minima depths).

\begin{figure}[t]
\centering
\includegraphics[trim = 16mm 0mm 8mm 10mm,clip,width=0.42\textwidth]{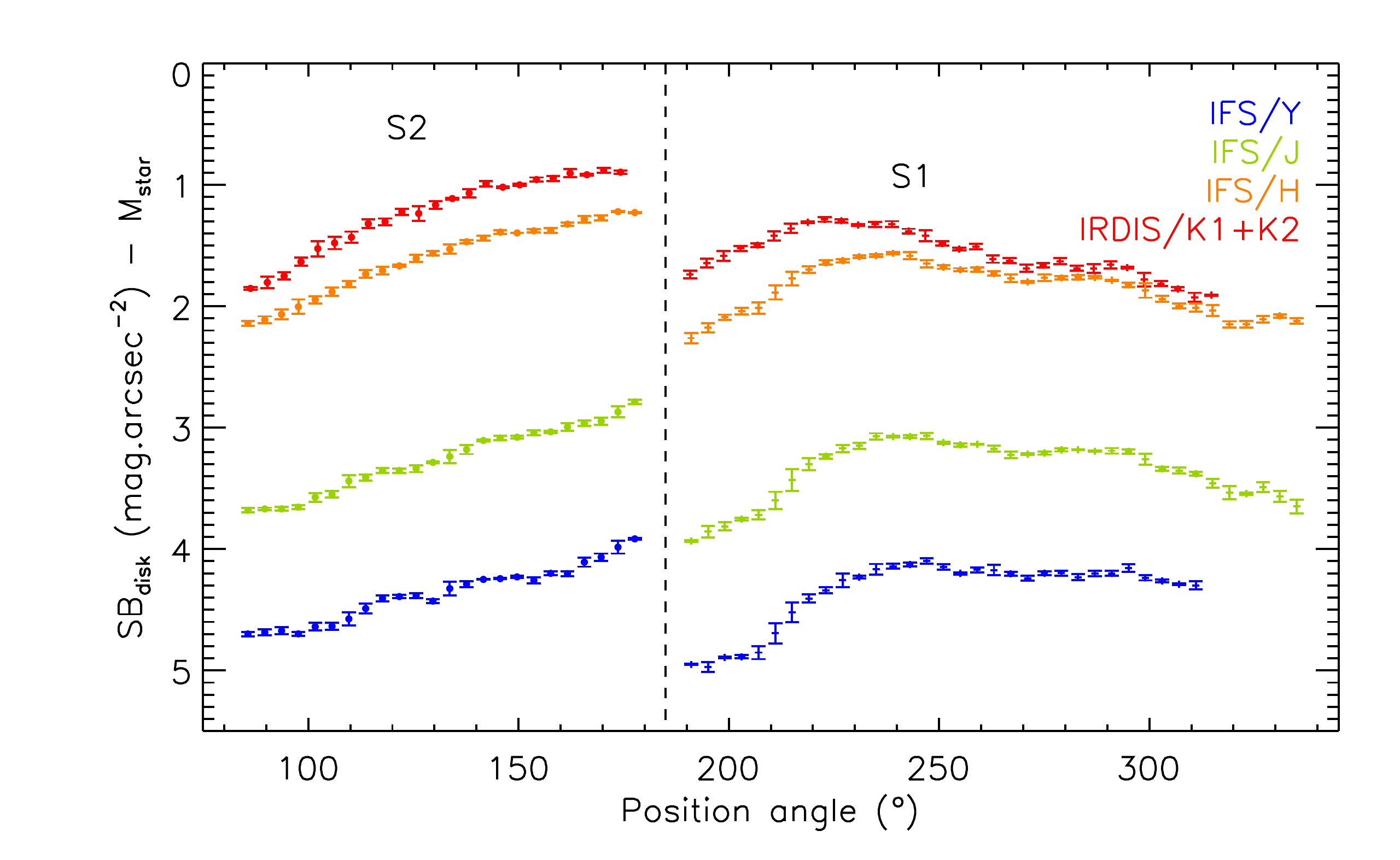}
\caption{{Intensity peaks of the spiral arms normalized to the unsaturated non-coronagraphic PSF intensity peak as a function of the {position angle} and the spectral band (see text). For the sake of \rev{clarity}, we considered a sampling of 4$^{\circ}$ for these measurements.}}
\label{fig:spiralpeak}
\end{figure}

\rev{Figure}~\ref{fig:spiralpeak} shows the surface brightness profiles along the spiral spines for the IFS and IRDIS bands. {We retrieve the} increasing trend of the disk surface brightness with the wavelength seen for the radial surface brightness profiles (Fig.~\ref{fig:brightnessprof}), especially for the IFS bands.

\begin{figure}[t]
\centering
\includegraphics[trim = 4mm 0mm 8mm 0mm,clip,width=0.42\textwidth]{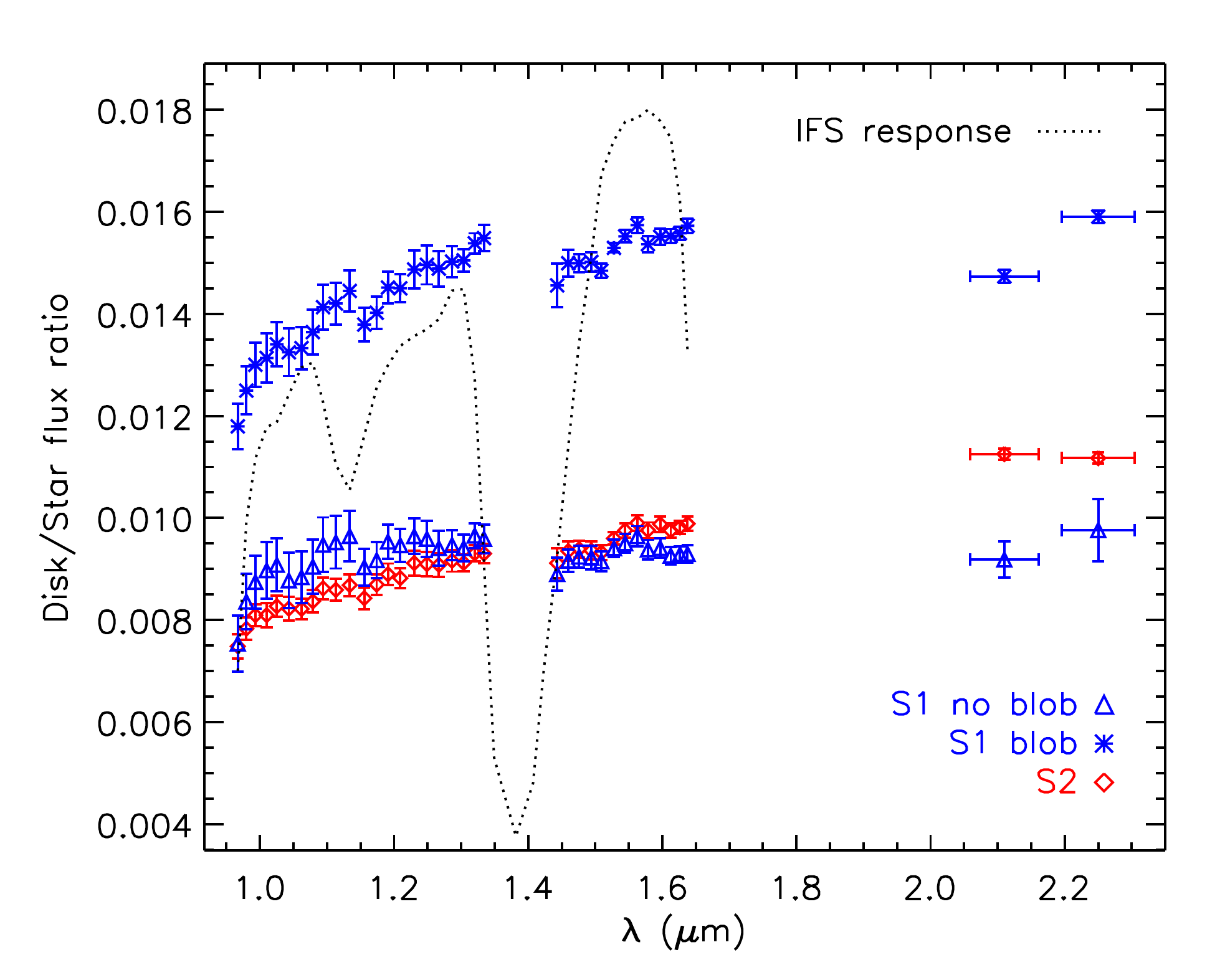}
\caption{{Reflectance spectra of the spiral arms covering the IFS and IRDIS passbands (see text). The horizontal error bars on the IRDIS measurements are the filter widths. The black dotted line shows the IFS instrument response normalized to 0.018. The spectral features at 1.13 and 1.38~$\muup$m are water telluric absorptions.}}
\label{fig:reflectancespectrum}
\end{figure}

Finally, we extracted from the deprojected and $r^2$-scaled RDI images the reflectance spectra integrated {on the S1 bright blob feature, the S1 regions outside its blob feature, and S2}. Figure~\ref{fig:reflectancespectrum} shows the spectra normalized to the {stellar spectrum measured in the unsaturated non-coronagraphic data}. The error bars on the measurements account for the errors in the photometric calibration {and uncertainties in the RDI subtraction}. We removed the spectral parts at $\sim$1.34--1.43~$\muup$m because of a strong water telluric absorption band which affects significantly the S/N of the corresponding images. \rev{The S1 region is brighter than S2,} as previously noted \citep{Garufi2013}, because of the contribution from {its bright blob feature}. {The spectra of the S1 blob feature and of S2 show a red slope in the $H$ band, whereas the spectrum of S1 outside its blob feature looks rather flat. The spectrum of S2 \rev{looks} flat in the $K$ band, where S1 appears slightly red. We also note spectral differences at the shortest \rev{wavelengths;} S2 and the S1 blob \rev{look red,} while the S1 regions outside its blob feature have a flatter spectrum. However, we stress that these \rev{S1} regions are poorly detected at wavelengths shorter than 1.13~$\muup$m because of the overlapping of remaining residuals from the AO correction radius ($R_{\rm{corr}}$\,=\,$N_{\rm{act}}$\,$\times$\,$\lambda/2D$, with $N_{\rm{act}}$ the number of actuators on one side of the deformable mirror). The scattering efficiency \rev{is higher} at longer wavelengths for S2, while for S1 it increases \rev{across} the $YJH$ bands \rev{and} then flattens in the $K$ band. This feature was also seen for the surface brightness profiles (Figs.~\ref{fig:brightnessprof} and \ref{fig:spiralpeak}). The spectral differences between S1 and S2 and within S1 could indicate that the surface dust grains might have different optical properties according to their location, which can be \rev{a} sign of different composition or size.} The SPHERE reflectance spectra of SAO~206462 look somewhat different from the SPHERE spectra of HR~4796A \citep{Milli2017}, which show an increasing slope at $\sim$1--1.1~$\muup$m \rev{and} then a plateau in the $J$ band.

\subsection{Summary}
The analysis of the SPHERE total intensity images confirm the red colors for the dust grains and the presence of the shadow lanes \rev{(though at lower significance)} found by \citet{Stolker2016} in SPHERE polarized-light images. They also indicate {sharp variations \rev{($\sim$40\%)} in the S1 width}, while the S2 width does not show strong variations. Finally, the reflectance spectra suggest {local spectral differences between S1 and S2 and within S1}, which might be a hint \rev{of} different composition or sizes of the dust grains. 

\begin{figure*}[t]
\centering
\includegraphics[trim = 8mm 0mm 5mm 0mm,clip,width=0.42\textwidth]{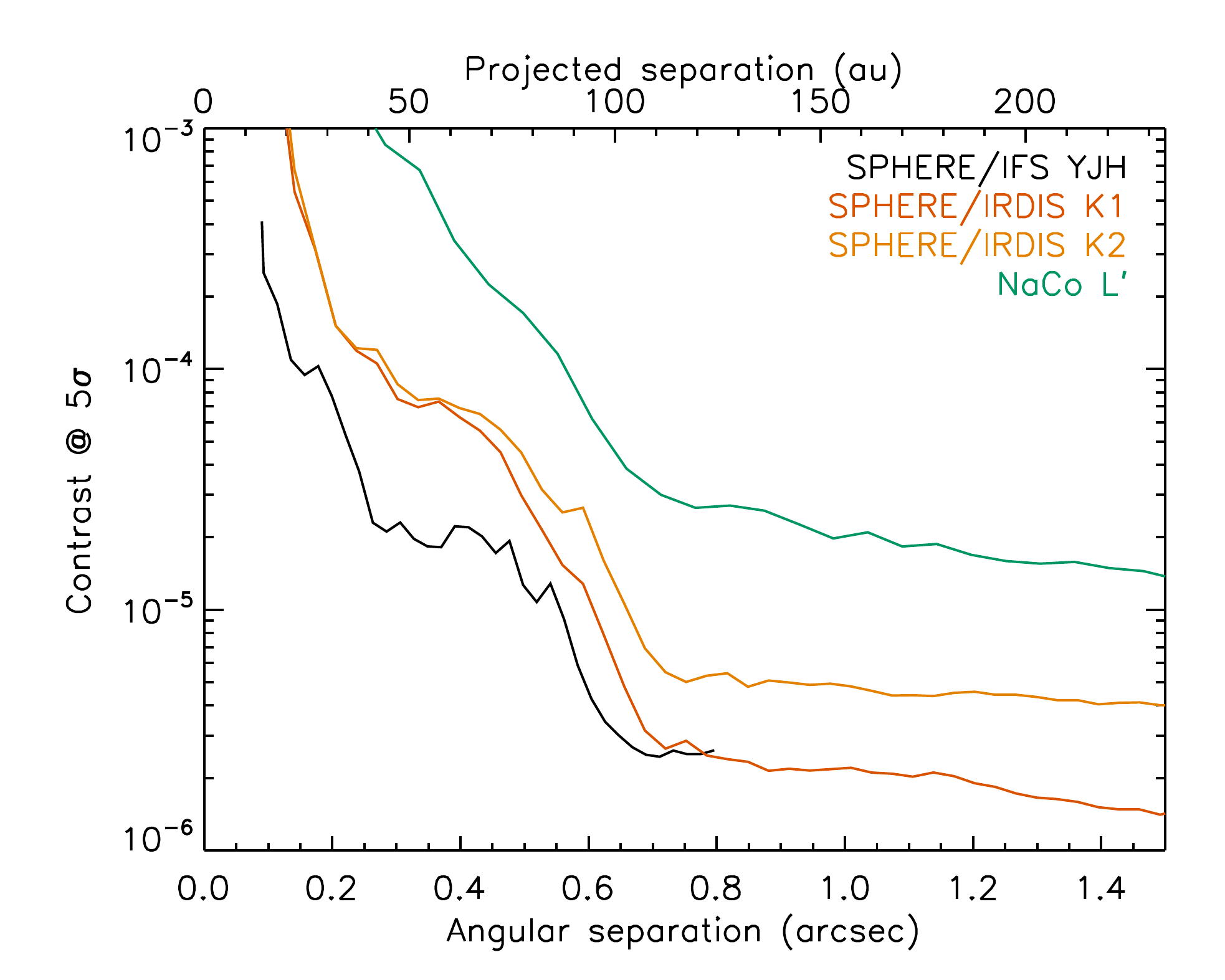}
\includegraphics[trim = 8mm 0mm 5mm 8mm,clip,width=0.42\textwidth]{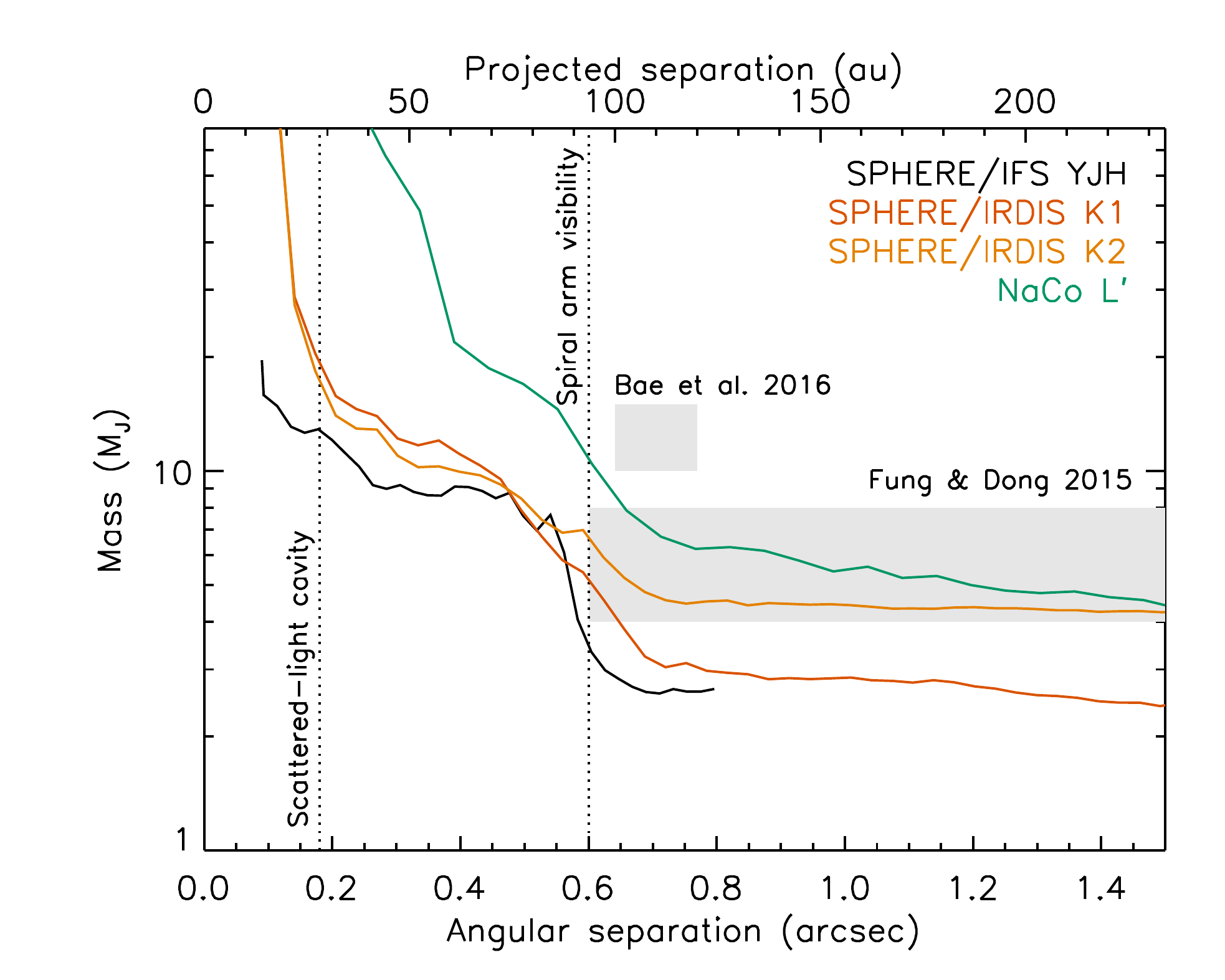}
\caption{{5$\sigma$ detection limits in contrast with respect to the star (\textit{left}) and in planet mass (\textit{right}). For the conversion of the SPHERE contrast limits, we used the {atmospheric and evolutionary models of \citet{Baraffe2015} and \citet{Baraffe2003}}, while for the NaCo contrast limits, we assumed the {models of \citet{2012RSPTA.370.2765A} and \citet{Baraffe2003}}. The locations of the outer edge of the scattered-light cavity and the outer visibility radius of the spiral arms are also shown \citep{Stolker2016}. {Greyed areas indicate theoretical planet predictions from \citet{Fung2015} and \citet{Bae2016}.}}}
\label{fig:detlimits}
\end{figure*}

\section{Detection limits on putative giant protoplanets}
\label{sec:detectionlimits}

\subsection{Archival NaCo/$L^{\prime}$ imaging data}
\label{sec:nacodata}
We reduced and analyzed archival NaCo/$L^{\prime}$ imaging data (ESO program 090.C-0443, P.I. Currie) using a custom pipeline for the reduction steps (cosmetics, frame registering) and the SPHERE consortium image processing tools (Sect.~\ref{sec:datareduc}). SAO~206462 \citep[$L$\,=\,4.89,][]{Coulson1995} was observed on 2013 March 24 UT around the meridian passage (airmass start/end 1.09--1.14) for $\sim$3.33\,h in pupil-tracking and saturated imaging modes. The star was regularly dithered between the bottom quadrants of the detector for sky background measurements. The observing conditions were good at the beginning of the sequence (DIMM seeing 0.5--0.7$''$ and coherence time $\sim$3\,ms)\rev{, but deteriorated} with time (DIMM seeing 1.1--1.4$''$ and coherence time $\sim$1.5\,ms at the end of the observations). \rev{Two hundred} data cubes of 160 co-adds of 0.25\,s were recorded covering a field of view rotation of 152.2$^{\circ}$. After binning the frames by groups of 100 and removing poor-quality frames based on the statistics of the total flux measured in an annulus of inner and outer radii 0.25$''$ and 0.5$''$, we were left with 226 images ($\sim$71\% of the complete sequence). The inner and outer radii of the annulus were chosen to exclude the regions dominated by the bright stellar PSF wings and the background noise. The unsaturated PSF was recorded several times at regular intervals during the sequence using the same individual integration time and number of co-adds with the ND\_LONG neutral density filter (transmission $\sim$1.78\%) for a total integration time of 480\,s. No point source was detected close to the star. We computed the TLOCI 5$\sigma$ detection limits using the same parameters as in Sect.~\ref{sec:datareduc} {and corrected them for the small sample statistics \citep{Mawet2014}.}

{The SPHERE and NaCo} contrast limits are shown in the left panel of Fig.~\ref{fig:detlimits}. The thermal background increases with the wavelength, hence the poorer contrast performance observed for IRDIS in the $K2$ band and NaCo in the $L^{\prime}$ band at large separations. The SPHERE contrasts are deeper than the NaCo contrasts by factors \rev{greater} than $\sim$10 in the speckle-limited regions ($\lesssim$0.6$''$), illustrating the improved performances of new dedicated high-contrast instruments like SPHERE and GPI.

\subsection{Comparison of the detection limits to planet predictions}

Several predictions for planet(s) shaping the disk of SAO~206462 have been proposed \citep{Muto2012, Garufi2013, Fung2015, Bae2016, vanderMarel2016b}. Using linear equations from the spiral density wave theory, \citet{Muto2012} suggested two planets with separations beyond $\sim$50~au by fitting independently the two spiral arms seen in Subaru/HiCIAO data and with masses of $\sim$0.5~$M_{\rm{J}}$ by using the amplitude of the spiral wave. \citet{Garufi2013} proposed that one planet of mass \rev{5--13~$M_{\rm{J}}$} located inside the cavity {at a separation of 17.5--20~au} could be responsible for the different cavity sizes measured for the small and large dust grains. \citet{Fung2015} presented scaling relations between the azimuthal separation of the primary and secondary arms and the planet-to-star mass ratio for a single companion on a circular orbit with a mass between Neptune mass and 16~$M_{\rm{J}}$ around a 1~$M_{\odot}$ star. They predicted \rev{with 30\% accuracy that a single putative planet responsible for both spiral features of SAO~206462 would have a mass of $\sim$6~$M_{\rm{J}}$}. \citet{Bae2016} presented dedicated hydrodynamical simulations of the SAO~206462 disk and proposed that both the bright scattered-light feature \citep{Garufi2013} and the dust emission peak \citep{Perez2014} seen for the southwestern spiral arm result from the interaction of the spiral arm with a vortex, although a vortex alone can account for the {S1 brightness peak}. They suggested that a 10--15~$M_{\rm{J}}$ planet may orbit at 100--120~au from the star. However, ALMA observations at two different frequencies seem to contradict a dust particle trapping scenario by a vortex \citep{Pinilla2015}. \citet{Stolker2016} performed new fitting of the spiral arms observed in SPHERE data and found a best-fit solution with two protoplanets located exterior to the spirals: $r_1$\,$\sim$\,168~au, $\theta_1$\,$\sim$\,52$^{\circ}$ and $r_2$\,$\sim$\,99~au, $\theta_2$\,$\sim$\,355$^{\circ}$. \citet{vanderMarel2016b} proposed that the features seen in thermal emission in ALMA data and the scattered-light spiral arms are produced by a single massive giant planet located inside the cavity at a separation of $\sim$30~au. {Recently, \citet{Dong2017} used the contrast of the spiral arms to predict a giant planet of $\sim$5--10~$M_{\rm{J}}$ at $\sim$100~au.}

We present \rev{the SPHERE and NaCo mass limits in the right panel of Fig.~\ref{fig:detlimits}}. We assumed a system age of 9~Myr \citep{Mueller2011} and the predictions of the {atmospheric and evolutionary ``hot-start'' models of \citet{Baraffe2015} and \citet{Baraffe2003} for the SPHERE limits and of \citet{2012RSPTA.370.2765A} and \citet{Baraffe2003} for the NaCo limits}, respectively\footnote{{The {\citet{Baraffe2015} atmospheric models} do not provide planet luminosities in the NaCo filters}.}. The detection limits are deeper with respect to the work of \citet{Vicente2011}, with the {exclusion of companions more massive than $\sim$14--12~$M_{\rm{J}}$ {in the range 0.12--0.2$''$ (19--31~au)}, $\sim$8~$M_{\rm{J}}$ beyond 0.5$''$ ($\sim$80~au) and $\sim$4~$M_{\rm{J}}$ beyond 0.6$''$ ($\sim$90~au)}. They do not allow us to strongly constrain the predictions of \citet{Muto2012}, \citet{Garufi2013}, and \citet{vanderMarel2016b}. They seem at odds with the predictions of a massive ($\sim$4--15~$M_{\rm{J}}$) giant planet in the outer disk by \citet{Fung2015}, \citet{Bae2016}{, and \citet{Dong2017}}. However, such models are expected to be degenerate. A less massive planet closer to the spiral arms should produce perturbations of similar amplitude \rev{to} a more massive planet farther out. Nevertheless, the pitch angle of the produced spiral arms changes with the location of the planet (which also corresponds to the launching point of the spiral wave) because of its dependency on the sound speed and disk temperature. For SAO~206462, the planet should still be more massive than $\sim$2~$M_{\rm{J}}$ according to the predictions of \citet{Fung2015} to account for the large pitch angle observed between the spiral arms ($\sim$130$^{\circ}$).

Estimating detection limits from high-contrast images in terms of planet mass depends on the assumptions for the system age and the mass-luminosity relation \citep[e.g.,][]{Marley2007, Spiegel2012, Mordasini2012a}. We note that for the upper age limit for the system of 16~Myr \citep{vanBoekel2005} the SPHERE {$K1$-band mass limits are $>$11--7~$M_{\rm{J}}$ beyond 0.5--0.7$''$} according to the {models of \citet{Baraffe2015, Baraffe2003}}. Hence, a massive giant planet formed following a ``hot-start'' formation mechanism would have been detected in the SPHERE data if located exterior to the spiral features. A massive giant planet in \rev{a} wide orbit such as predicted by \citet{Fung2015}, {\citet{Bae2016}, and \citet{Dong2017}} would still be compatible with the SPHERE constraints if formed according to a ``warm-start'' scenario with low initial entropy \citep[for e.g., a 10-$M_{\rm J}$ planet with age 16~Myr and initial entropy {below 10~$k_B$/baryon} could not be excluded assuming the models of][]{Spiegel2012}.

{Another possible planet scenario for SAO~206462 would be a low-mass protoplanet embedded in the disk and surrounded by a circumplanetary disk. This scenario has been proposed for planet companions/candidates in the transitional disks of HD~100546 \citep{Quanz2013a, Quanz2015, Currie2015} and HD~169142 \citep{Biller2014, Reggiani2014}. The mass-luminosity relations of standard evolutionary models \citep[e.g.,][]{Baraffe2003} assume giant planets without circumplanetary disks. A giant planet surrounded by a circumplanetary disk could be significantly brighter than the same planet without a disk, hence the mass-luminosity relations of standard evolutionary models could be pessimistic. Therefore, the mass limits shown in Fig.~\ref{fig:detlimits} could be upper mass limits for a given system age. However, a lower-mass giant planet with a circumplanetary disk would exert weaker gravitational perturbations on the disk than a more massive giant planet without a disk and produce a different spiral shape (e.g., smaller opening angle, smaller spiral contrast). \rev{Modeling} studies of disk spirals suggested that lower planet mass limits could be estimated using an observed disk spiral pattern \citep[][ see previous paragraph]{./figs/fung2015}.}

\subsection{Summary}

The SPHERE near-IR detection limits improve significantly the constraints on putative planets in the SAO~206462 disk with respect to the study of \citet{Vicente2011} and the detection limits measured in archival NaCo thermal IR images. The SPHERE data exclude massive giant planets ($>$3~$M_{\rm J}$ assuming a ``hot-start'' scenario and a system age of 9~Myr) exterior to the spiral arms, which may rule out a few recent {predictions based on spiral modelling \citep{Fung2015, Bae2016, Dong2017}}. However, we cannot exclude low-mass giant planets in the outer disk and/or giant planets inside the scattered-light cavity, as predicted in other studies \citep{Muto2012, Garufi2013, vanderMarel2016b}. These planets could still account for the morphology of the SAO~206462 spirals.

\section{\rev{Effect} of protoplanetary dust on the detectability of embedded planets}
\label{sec:dustopacity}

The detection of planets embedded in protoplanetary disks is expected to be hampered by the dust grains residing between the disk midplane and surface \citep[e.g.,][]{Quanz2015}, hence Fig.~\ref{fig:detlimits} likely provides limits of detection \rev{that are too optimistic}. However, a massive planet will open a (partial) gap \citep[e.g.,][]{Crida2006, Malik2015} leading to a smaller attenuation of a planet's thermal emission. Since the dust opacity decreases with the wavelength, the SPHERE/$YJHK$ detection limits are expected to be more affected than the NaCo/$L^{\prime}$ detection limits.

\subsection{Methods and assumptions}

We determined the minimum masses of embedded planets that would be detectable given the derived IRDIS/$K1$ and NaCo/$L^{\prime}$ contrast limits (Fig.~\ref{fig:detlimits}) by assuming a simple disk structure and considering {theoretical mass-luminosity relations for giant planets without circumplanetary disk from \citet{Baraffe2015, Baraffe2003}} at the age of the SAO~206462 system. We used DIANA standard dust opacities \citep{Woitke2016} in the $K1$ and $L^{\prime}$ bands of 330 and 250~cm$^2$/g, respectively, and a surface density profile inversely proportional to the distance to the star. We assumed a total dust mass of $2\times 10^{-4}$~$M_{\odot}$, an {inner and outer disk radius of 28 and 300~au} \citep{Carmona2014} and a zero surface density within the {scattered light cavity (28~au)}. Since the disk is observed close to face-on \citep{vanderMarel2016a}, we have neglected projection effects.

\begin{figure}[t]
\centering
\includegraphics[width=0.44\textwidth]{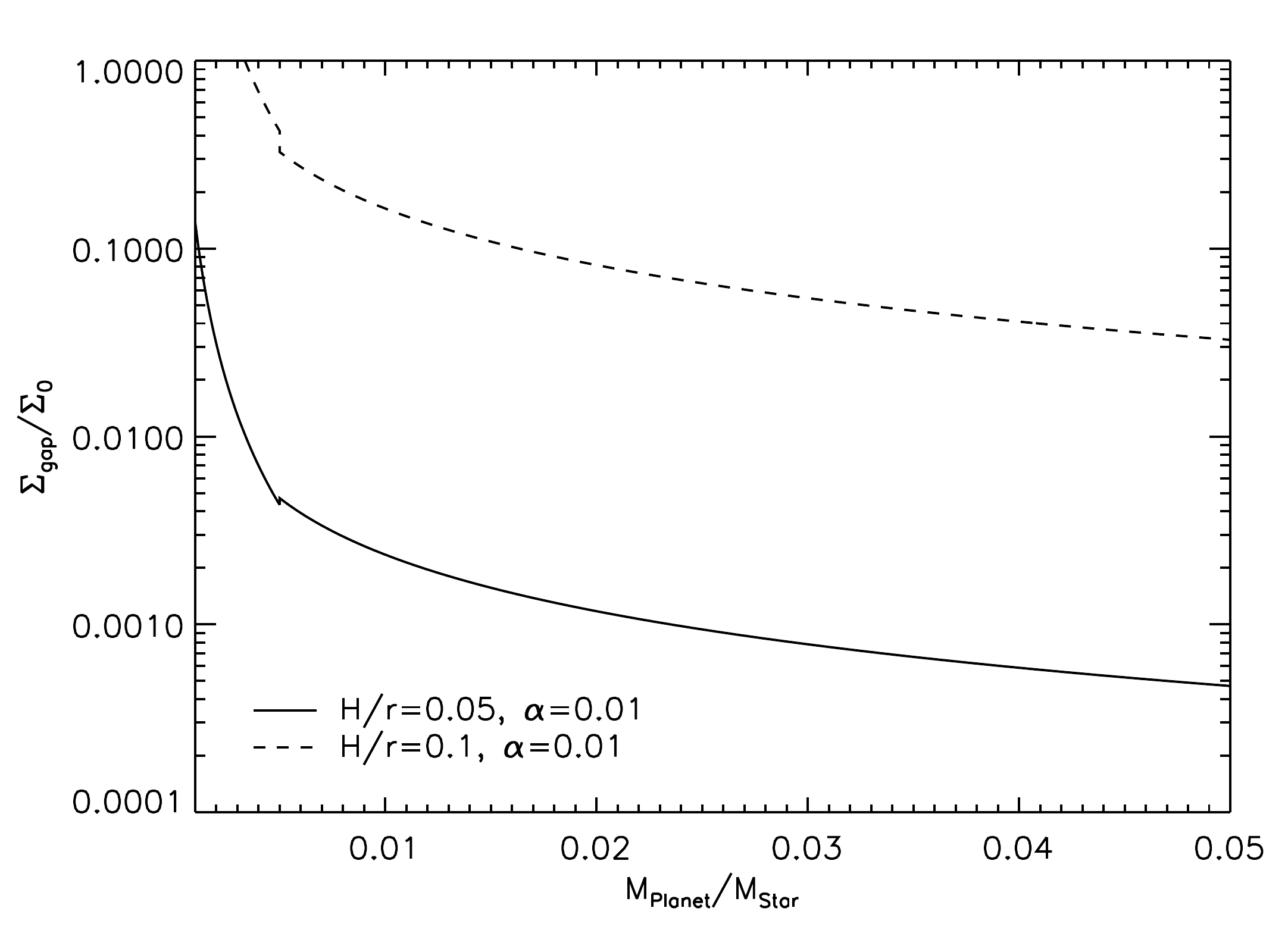}
\caption{Depth of gaps opened by putative giant planets in the disk of SAO~206462 as a function of the planet-to-star mass ratio computed from {the empirical relations in \citet{Fung2014}} for the two cases considered in this paper (see text). \rev{For the pessimistic case (dashed curve) and a planet-to-star mass ratio below $\sim$3.5$\times$10$^{-3}$ (i.e., planet mass below $\sim$6~$M_{\rm J}$), a planet cannot open a gap in the disk, hence the dust column density above the planet is not reduced.}}
\label{fig:planetgapdepths}
\end{figure}

Then, we take a planet gap depth into account by following the empirical relations of \rev{\citet{Fung2014}}
\begin{eqnarray}
\frac{\Sigma_{\rm p}}{\Sigma_0} & = & 0.14\left( \frac{q}{10^{-3}} \right)^{-2.16} \left(\frac{\alpha}{10^{-2}} \right)^{1.41} \left( \frac{H/r}{0.05} \right)^{6.61} \label{eq:fung1}\\
 & & \mathrm{for}~10^{-4} \leq q \leq 5\times10^{-3},  \nonumber \\
\frac{\Sigma_{\rm p}}{\Sigma_0} & = & 4\times10^{-3}\left( \frac{q}{5\times10^{-3}} \right)^{-1.00} \left(\frac{\alpha}{10^{-2}} \right)^{1.26} \left( \frac{H/r}{0.05} \right)^{6.12} \label{eq:fung2} \\
& & \mathrm{for}~5\times10^{-3} \leq q \leq 10^{-2} \nonumber,
\end{eqnarray}
where $\Sigma_{\rm p}$ is the surface density at the orbital radius of the planet, $\Sigma_0$ the initial surface density before a gap was formed, $q$ the planet-to-star mass ratio, $\alpha$ the dimensionless Shakura–Sunyaev viscosity parameter and $H/r$ the disk aspect ratio. Eqs.~(\ref{eq:fung1}) and (\ref{eq:fung2}) are strongly dependent on the assumed properties for the disk and the dust grains, which are not well constrained by observations and/or \rev{modeling} of the disk of SAO~206462. We had to make assumptions on several of these parameters for our analysis in the remainder of this section. We note that our quantitative results are highly sensitive to our parameter choice. \citet{Fung2014} investigated the morphology of planet gaps up to mass ratios of 10$^{-2}$ (i.e., planet masses below 17~$M_{\rm J}$ in the case of SAO~206462), while the {NaCo and IRDIS detection limits} probe planet-to-star mass ratios above this limit at close-in separations beyond the scattered-light cavity (Fig.~\ref{fig:detlimits}). {For these cases, we simply used Eq.~(\ref{eq:fung2})}. For the disk aspect ratio and viscosity, we considered two extreme cases shown in Fig.~\ref{fig:planetgapdepths}: (1) deep planet gaps by assuming $H/r=0.05$ and $\alpha=10^{-2}$ (\rev{for} this value of $H/r$, the assumed value for $\alpha$ has little effect on the derived results) and (2) shallow planet gaps corresponding to $H/r=0.1$ and $\alpha=10^{-2}$. The vertical optical depth to the disk midplane is now given by $\tau = \kappa \Sigma/2$ , with $\kappa$ the dust opacity, and the attenuation of a planet's thermal emission is a factor $e^{-\tau}$.

\subsection{Mass limits vs. disk aspect ratio and viscosity}

\begin{figure}[t]
\centering
\includegraphics[width=0.46\textwidth]{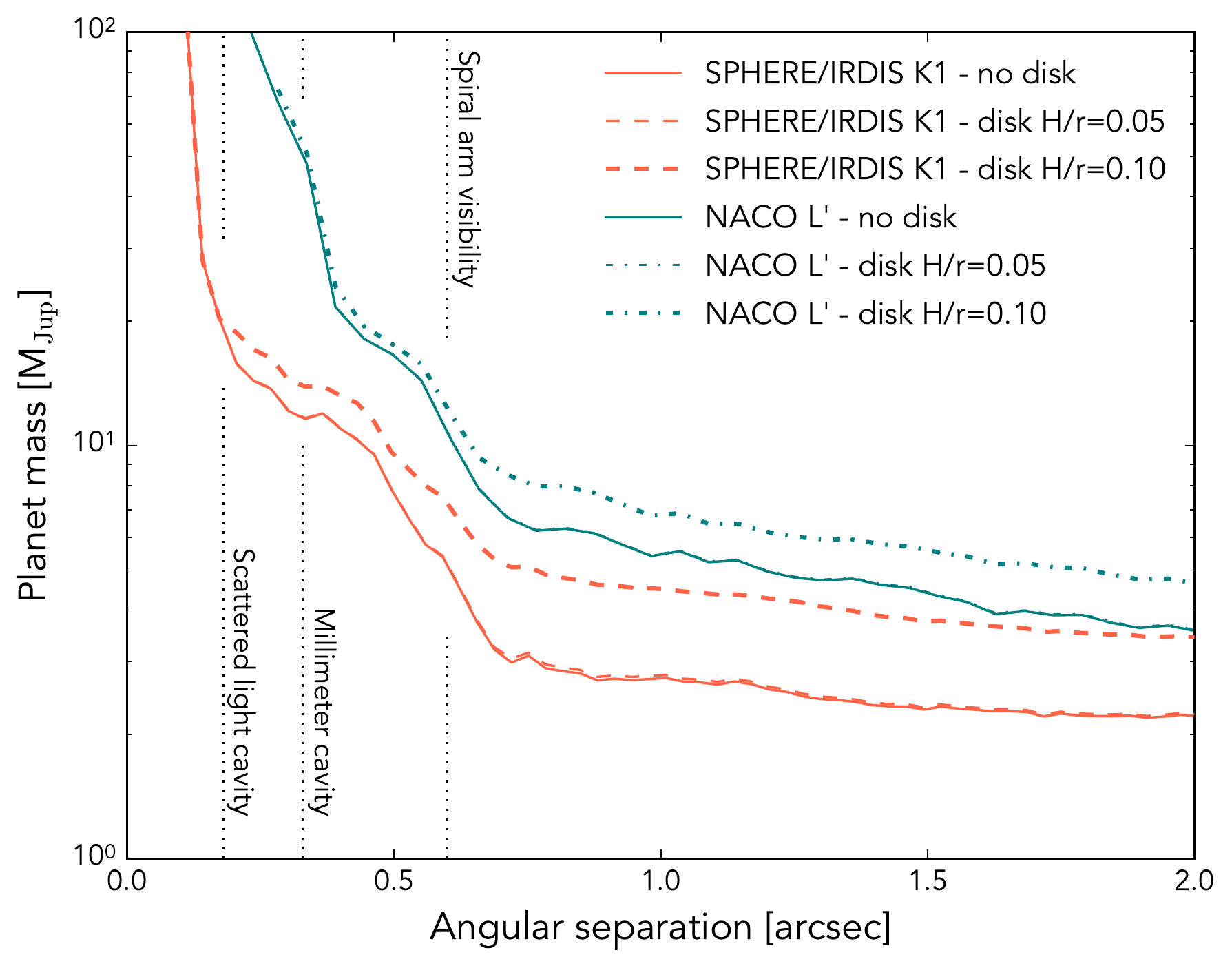}
\caption{{Planet mass detection limits (5$\sigma$) for the IRDIS $K1$-band and NaCo $L^{\prime}$-band images, both with the effect of attenuation of a planet emitted flux by protoplanetary dust (dashed and dash-dotted curves, see text) compared to the case where the attenuation by protoplanetary dust is neglected (solid curves, see Fig.~\ref{fig:detlimits}).} }
\label{fig:planetmassesopacity}
\end{figure}

Figure~\ref{fig:planetmassesopacity} shows the IRDIS/$K1$ and NaCo/$L^{\prime}$ planet mass limits for the two cases of protoplanetary dust attenuation compared to the limits for which the attenuation has been neglected. We note that for a disk aspect ratio of 0.05, the attenuation by the dust is negligible even for a \rev{high} viscosity because the disk properties are more favorable to the formation of deep gaps (Fig.~\ref{fig:planetgapdepths}). For a disk aspect ratio of 0.1 combined with a \rev{high} viscosity, the effect is significant at all separations beyond the scattered-light cavity (for $\alpha=10^{-3}$, the mass limits are degraded by less than $\sim$0.5~$M_{\rm J}$). This is expected because the planet gaps are shallower (Fig.~\ref{fig:planetgapdepths}). In particular, a planet less massive than $\sim$6~$M_{\rm J}$ cannot open a gap so that the dust column density above the planet is not reduced for this mass range. We also note that the IRDIS/$K1$ mass limits are deeper than the NaCo/$L^{\prime}$ limits. The decreased sensitivity to the protoplanetary dust opacity in the $L^{\prime}$ band does not compensate for the poorer contrasts provided by the NaCo observations. Higher contrast observations in the $L^{\prime}$ band by $\sim$1~mag could compete with our SPHERE data.

\subsection{Detectability of gaps opened by giant embedded planets}

\rev{In this section we test} the detectability of the gaps opened by putative embedded planets with the masses probed by SPHERE in Fig.~\ref{fig:planetmassesopacity}. In particular, we want to investigate the detectability of a planet gap beyond the spiral arms because several authors \rev{have} suggested that a single massive ($\sim$4--15~$M_{\rm J}$) planet located in these regions might be responsible for both spiral features \citep{Dong2015,Zhu2015,Fung2015, Bae2016}. Our SPHERE/IRDIS observations seem to reject the high-mass tail of these predictions, since the mass limits beyond the spiral arms are $\sim$5~$M_{\rm J}$ (see red dashed curve for $H/r$\,=\,0.05 and $\alpha$\,=\,0.01 in Fig.~\ref{fig:planetmassesopacity}). However, a giant planet of $\sim$4~$M_{\rm J}$ \rev{close} to the spiral arms ($\sim$0.6--0.65$''$ if $H/r$\,=\,0.05 and $\alpha$\,=\,0.01) would still be compatible with the detection limits \rev{and} could account for the spiral morphology.

\begin{figure}[t]
\centering
\includegraphics[trim = 2mm 0mm 2mm 0mm, clip, width=0.47\textwidth]{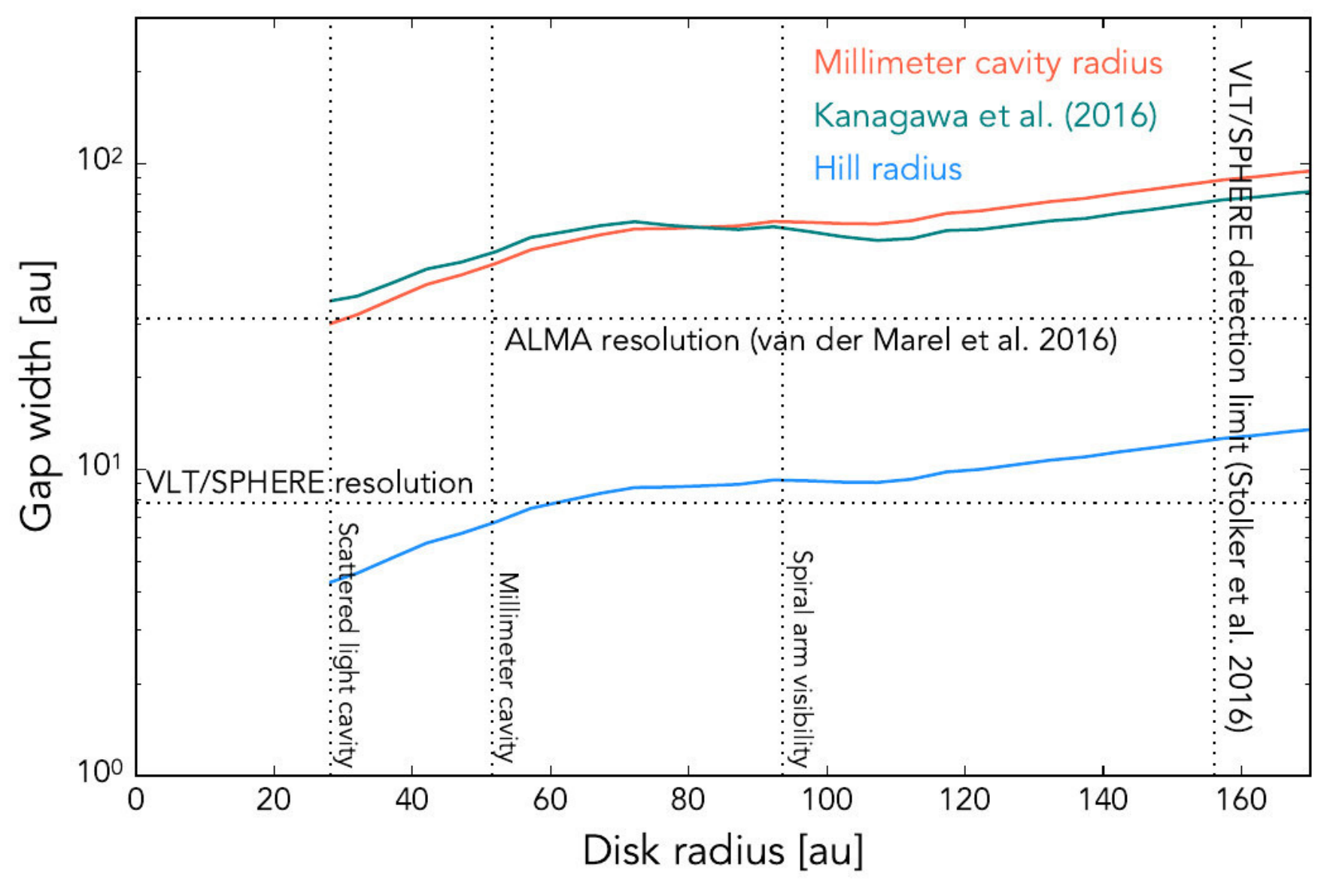}
\caption{{Gap widths in the gas and large dust grains for the $K1$-band planet mass limits, corrected for the attenuation by the dust (see Fig.~\ref{fig:planetmassesopacity}). The gas gap widths are assumed to be valid for the small dust grains, which are probed by the SPHERE data. The Hill radius gives a first-order estimate of a gap width in the gas {(blue)}. {We also considered the empirical relation from the hydrodynamical simulations of \citet{Kanagawa2016} (green curve, see text for details)}. The cavity radius in the large dust grains (red) is assumed to be $\sim$7~$R_{\rm H}$ following the work of \citet{Pinilla2012}.}} 
\label{fig:planetgaps}
\end{figure}

For this analysis, we considered {the case $H/r=0.05$ and $\alpha=10^{-2}$}. This case allows for the detection of lower-mass giant planets in the disk, hence the widths of the gaps produced by these planets may be considered as lower limits. We show in Fig.~\ref{fig:planetgaps} the width of the planet gaps as a function of the orbital radius\footnote{{The curves were cut to separations larger than the scattered-light cavity radius because we assumed zero surface density for the dust inside the cavity.}}. The Hill radius $R_{\rm H}$ is a first-order estimate of the gap width in the gas and can be applied to gaps in the small dust grains (traced in scattered light) assuming that they are coupled to the gas. However, this is a lower limit since the gas gap width can be as large as 5~$R_{\rm H}$ \citep{Dodson2011}. Therefore, we also used the {empirical formula from \citet{Kanagawa2016}, derived from 2D hydrodynamical simulations with planets of mass ratios 10$^{-4}$--2\,$\times$\,10$^{-3}$ (i.e., $\sim$0.2--4~$M_{\rm J}$ in the case of SAO~206462)}. When comparing the latter gap widths to the SPHERE resolution, we find that the gaps are significantly larger, meaning that a massive planet that is located beyond the spiral arms might have opened a detectable gap. Interestingly, the disk was marginally detected in scattered light beyond the spiral arms up to 1$''$ by \citet{Stolker2016} without any indication of a gap in that region. \rev{Given that the disk is not exactly seen face-on, we finally note that we} cannot rule out shadowing of the gaps by their outer/inner wall because of the uncertainties in the disk scale height.

However, scattered-light observations probe the disk surface so the detectability of a planet gap depends on the disk aspect ratio which, for a flaring disk, increases toward larger disk radii \citep{Crida2006}. Sub-mm data, like those provided by ALMA, probe the dust grains located in the disk midplane, \rev{and} hence are more relevant for such an analysis. \rev{Figure}~\ref{fig:planetgaps} shows the approximate gap width (7~$R_{\rm H}$) in the large, mm-sized grains as a result of dust trapping by a gas giant planet of {mass ratio 10$^{-3}$--3$\times$10$^{-3}$ in a pressure maximum} \citep{Pinilla2012}. The gap width is also larger than the angular resolution of recent ALMA observations \citep{vanderMarel2016b}. However, the disk appears to be compact in both $^{13}$CO and dust continuum \citep[e.g.,][]{vanderMarel2016a} with only a marginal detection of the disk beyond 0.5$''$. This feature might indicate that the disk is truncated by a companion located beyond the spiral arms, which would support the interpretation by \citet{Dong2015}. On the other hand, the low S/N of the sub-mm detection beyond 0.5$''$ might have been limited by the sensitivity of the ALMA observations.

\subsection{Summary}

{Since the disk dust may attenuate the signal of an embedded planet and lead to an underestimation of the mass detection limits, we estimated the amplitude of this potential effect for the SAO~206462 disk.} The analysis is in fact strongly dependent on the assumptions on the disk properties, especially its aspect ratio and viscosity. By considering two extreme cases for the planet gap depths, {we showed that the mass limits in Fig.~\ref{fig:detlimits} might be underestimated by up to $\sim$2~$M_{\rm J}$ exterior to the spiral arms}. Then, we analyzed the detectability of gaps opened by planets with masses compatible with the detection limits in the {case of deep planet gaps}. The comparison of the gap widths for small dust grains to the SPHERE resolution suggests that such gaps might have been resolved outside the scattered-light cavity. Nevertheless, the sensitivity of the scattered-light data to planet gaps is potentially hampered by the disk scale height, which is poorly constrained for SAO~206462. Therefore, we considered ALMA data and found that planet gaps in the mm-sized grains could also have been resolved exterior to the cavity measured at these wavelengths. While current ALMA data exclude such gaps at separations {up to 80~au}, they cannot exclude them at larger separations because of their limited sensitivity. {We note that very little dust/gas is detected outside of the spiral arms \rev{where} putative planets are predicted to reside, so both ALMA and SPHERE data do not strongly constrain the presence of putative planet gaps.}

\section{Conclusions}

We presented new high-resolution and high-contrast observations with the instrument VLT/SPHERE of the transitional disk of SAO~206462 to search for giant embedded planets responsible for its spiral features and, in \rev{the} case of non-detection, constrain recent predictions for such planets. We also exploited the spatial and spectral resolutions offered by SPHERE to analyze the colors of the disk spiral arms and to search for spatial variations in the spectral properties of the dust grains at the surface of the disk. The main results of the analysis are \rev{as follows}:\\

\textit{Search for giant protoplanets and constraints on predictions}
\begin{enumerate}
\item The IRDIS data \rev{shows} a point source at a separation of $\sim$5$''$ in addition to the known background stellar binary at the same separation range. From the IRDIS colors and the IRDIS and HST/STIS astrometry of the point source, we \rev{classify} it as a background star.

\item {Considering that the disk dust may attenuate the signal of a planet embedded within a disk, we evaluated that the mass detection limits computed from the contrast achieved in the SPHERE data may be underestimated by up to $\sim$2~$M_{\rm J}$ at separations exterior to the spiral arms}.

\item The SPHERE detection limits allowed us to exclude the presence of planets more massive than $\sim$5--7~$M_{\rm J}$ beyond the spiral arms, which provides new constraints for the hydrodynamical \rev{modeling} of the disk spiral arms. The non-detection of planet gaps up to {separations of $\sim$80~au} in ALMA data do not favor the presence of massive giant planets in the outer disk close to the spiral arms, although we cannot exclude that the non-detection might be related to a lack of resolution of the instrument. {However, very little disk signal is detected in the ALMA data outside the spiral arms, so the observational constraints on the presence of planet gaps are loose.} A giant planet of $\sim$4--6~$M_{\rm J}$ in the outer disk would still be compatible with the SPHERE data \rev{and} can account for the spiral morphology. Another possibility would be that the spiral arms are driven by one or more giant planets located inside the scattered-light cavity.

\item Although NaCo/$L^{\prime}$-band data are less sensitive to protoplanetary dust opacity than SPHERE near-IR observations, the poorer contrasts achieved in archival NaCo data in this band make the mass limits shallower than SPHERE/IRDIS limits in the $K1$ band.

\end{enumerate}

\textit{Imaging and spectrophotometry of the disk features}
\begin{enumerate}
\item We confirmed the shadow lanes detected in SPHERE polarized data \citep{Stolker2016}, although at lower significance.
\item We found sharp {variations of $\sim$40\% for the width of spiral arm S1}, while the width of spiral arm S2 does not exhibit strong variations.
\item We found \rev{a factor of $\sim$10 increase} for the disk surface brightness from the $Y$ to the $K$ bands, hinting at an increasing scattering efficiency of the dust grains at the surface of the disk. This could be explained by the presence of $\muup$m-sized dust aggregates.
\item The reflectance spectra suggest small spectral variations for the dust properties between {spiral arms S1 and S2 and between the bright blob feature and the other parts of S1}.

\end{enumerate}

\rev{In order to perform further searches} for protoplanets in the SAO~206462 disk, dedicated SPHERE observations might \rev{increase} the contrast/mass constraints beyond the disk spirals. Another method of interest would be to use H$\alpha$ imaging using SPHERE or MagAO given the system young age and that young protoplanets may still accrete. In addition, sparse aperture masking observations with 8--10~m ground-based telescopes will allow \rev{the scattered-light cavity for substellar companions to be probed} as close as $\sim$50--200~mas {($\sim$8--31~au)}, well within the angular resolution provided by high-contrast coronagraphic imagers on the same class of telescopes, but at limited contrasts of $\sim$10$^{2}$--10$^{3}$ {\citep[$\sim$30--300~$M_{\rm{J}}$ in the $L^{\prime}$ band for the age of SAO~206462 according to the predictions of][]{2012RSPTA.370.2765A, Baraffe2003}}. We note that putative embedded planets surrounded by circumplanetary disks could be detected with these instruments down to significantly lower masses. In the near future, JWST imaging will probe for very low-mass planets outside the spiral arms, while high-contrast imagers on 30--40~m telescopes will be required to search the scattered-light cavity for planetary-mass companions.

New promising possibilities for disk analyses are now offered by the high angular resolution combined with the spectral and/or polarimetric capabilities of the high-contrast imaging instruments SPHERE, GPI, and CHARIS. In particular, the spectra of distinct parts of a disk can be compared to search for differences indicating the presence of dust grain populations with different properties (e.g., scattering efficiency, size). 

Finally, we outline the complementarity of scattered-light and thermal emission information\rev{, such} as those brought by SPHERE and ALMA to constrain the presence of giant protoplanets in transitional disks. More multiwavelength analyses of such targets will help to better understand the relations between some features seen in these disks (spirals, gaps, rings) and putative giant embedded planets. These analyses will require further multiwavelength efforts. 

\begin{acknowledgements}
     The authors thank the ESO Paranal Staff for support in conducting the observations and Philippe Delorme and Eric Lagadec (SPHERE Data Center) for their help with the data reduction. We thank Myriam Benisty, Philippe Th\'ebault, and an anonymous referee for helpful comments. A.-L.M., S.M., S.D., R.G., and D.M. acknowledge support from the ``Progetti Premiali'' funding scheme of the Italian Ministry of Education, University, and Research. We acknowledge financial support from the Programme National de Planétologie (PNP) and the Programme National de Physique Stellaire (PNPS) of CNRS-INSU. This work has also been supported by a grant from the French Labex OSUG@2020 (Investissements d'avenir -- ANR10 LABX56). M.R.M and S.P.Q. acknowledge the financial support of the SNSF. The project is supported by CNRS, by the Agence Nationale de la Recherche (ANR-14-CE33-0018). This work has made use of the SPHERE Data Centre, jointly operated by OSUG/IPAG (Grenoble), PYTHEAS/LAM/CeSAM (Marseille), OCA/Lagrange (Nice) and Observatoire de Paris/LESIA (Paris). This research made use of the SIMBAD database and the VizieR Catalogue access tool, both operated at the CDS, Strasbourg, France. The original description of the VizieR service was published in Ochsenbein et al. (2000, A\&AS 143, 23). SPHERE is an instrument designed and built by a consortium consisting of IPAG (Grenoble, France), MPIA (Heidelberg, Germany), LAM (Marseille, France), LESIA (Paris, France), Laboratoire Lagrange (Nice, France), INAF -- Osservatorio di Padova (Italy), Observatoire astronomique de l'Universit\'e de Gen\`eve (Switzerland), ETH Zurich (Switzerland), NOVA (Netherlands), ONERA (France), and ASTRON (Netherlands), in collaboration with ESO. SPHERE was funded by ESO, with additional contributions from the CNRS (France), MPIA (Germany), INAF (Italy), FINES (Switzerland), and NOVA (Netherlands). SPHERE also received funding from the European Commission Sixth and Seventh Framework Programs as part of the Optical Infrared Coordination Network for Astronomy (OPTICON) under grant number RII3-Ct-2004-001566 for FP6 (2004--2008), grant number 226604 for FP7 (2009--2012), and grant number 312430 for FP7 (2013--2016).
\end{acknowledgements}

%
   \bibliographystyle{aa} 
   \bibliography{biblio} 
%

\begin{appendix} 

\section{IFS individual images}
\label{sec:ifsimchannels}

\begin{figure*}[t]
\centering
\includegraphics[trim = 0mm 0mm 0mm 10mm,clip,width=0.175\textwidth]{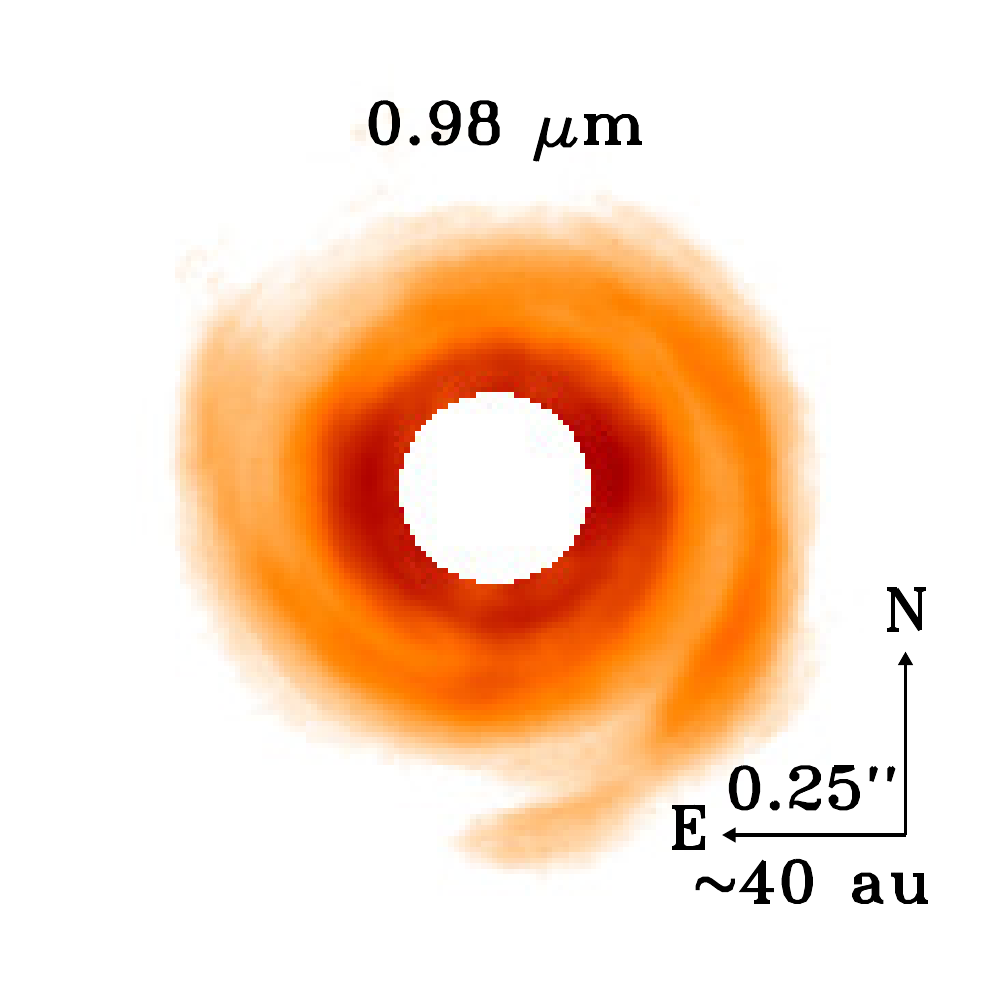}
\includegraphics[trim = 0mm 0mm 0mm 10mm,clip,width=0.175\textwidth]{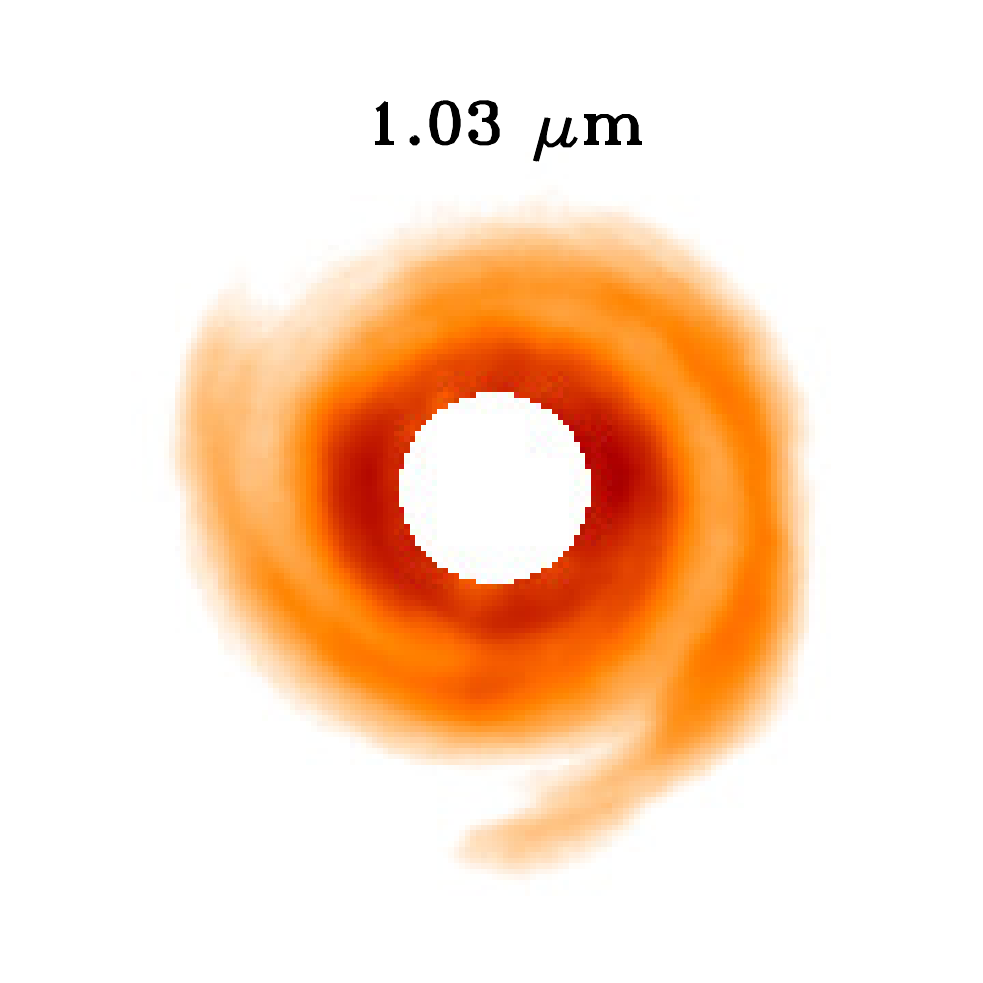}
\includegraphics[trim = 0mm 0mm 0mm 10mm,clip,width=0.175\textwidth]{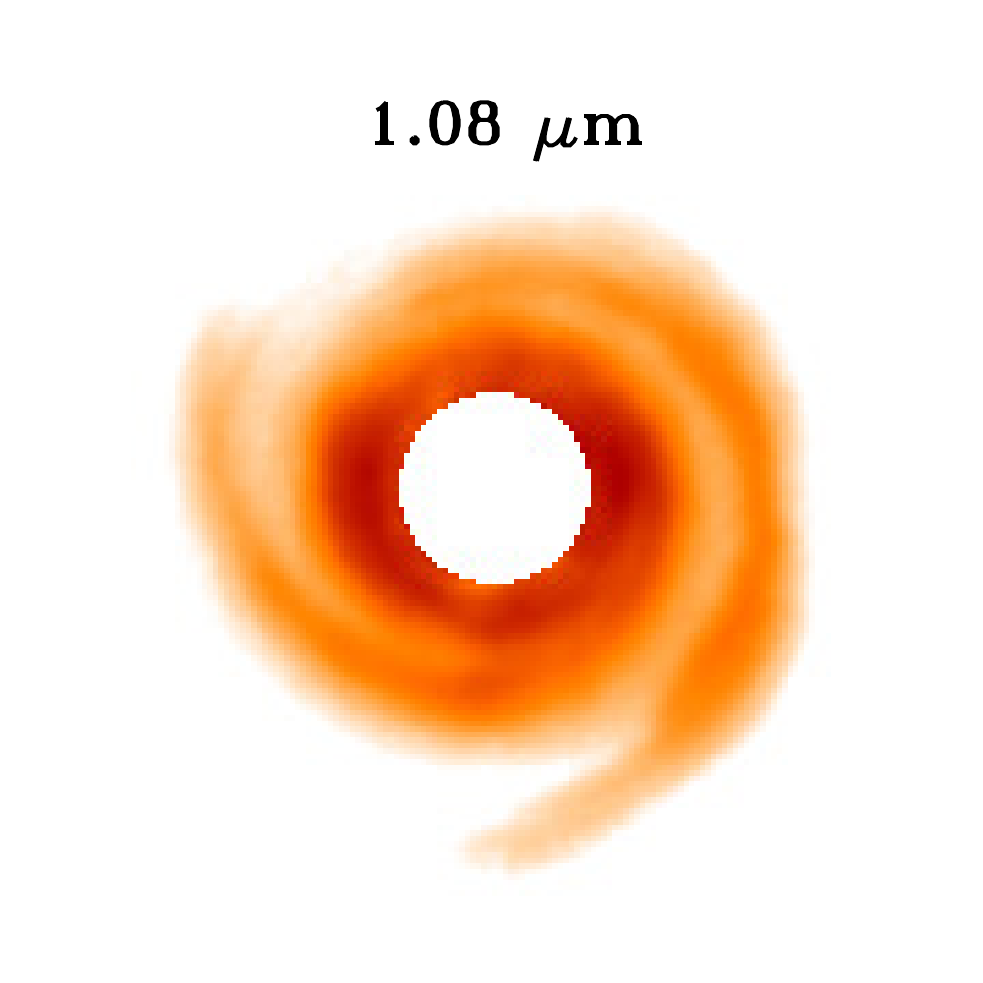}
\includegraphics[trim = 0mm 0mm 0mm 10mm,clip,width=0.175\textwidth]{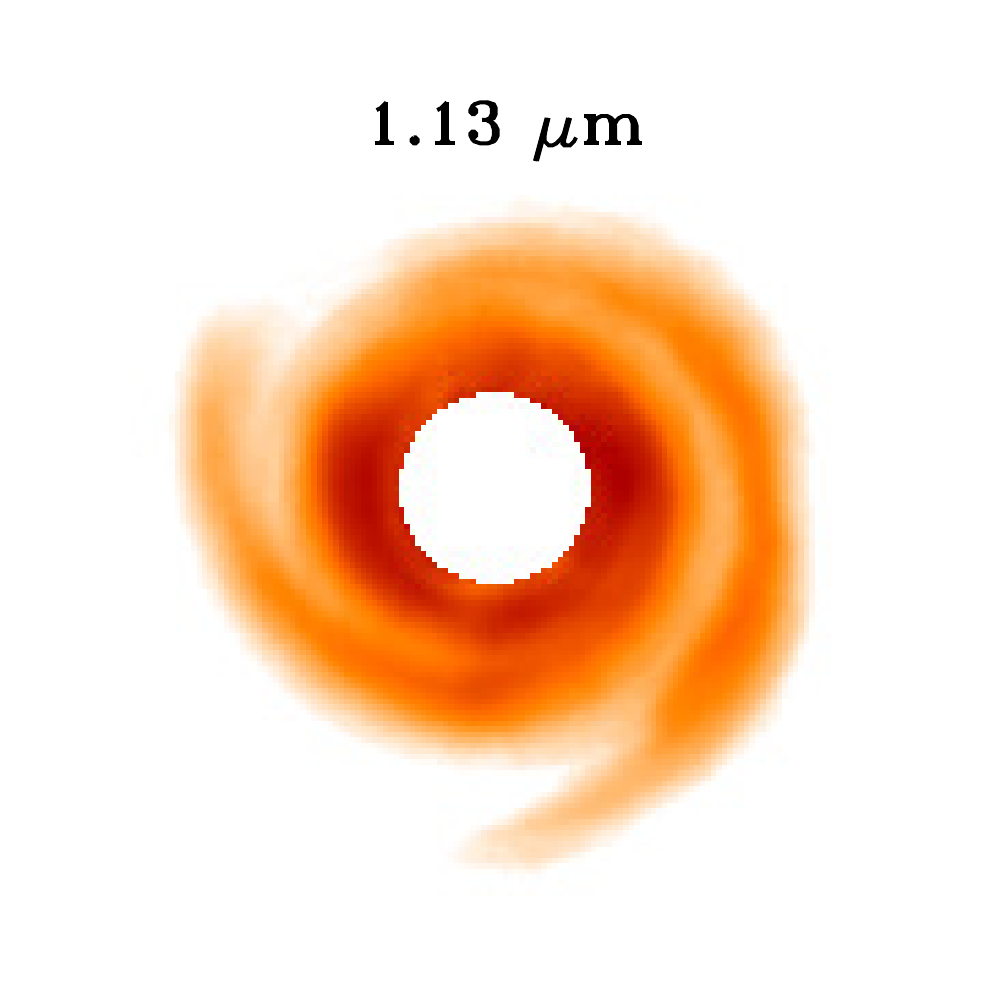}
\includegraphics[trim = 0mm 0mm 0mm 10mm,clip,width=0.175\textwidth]{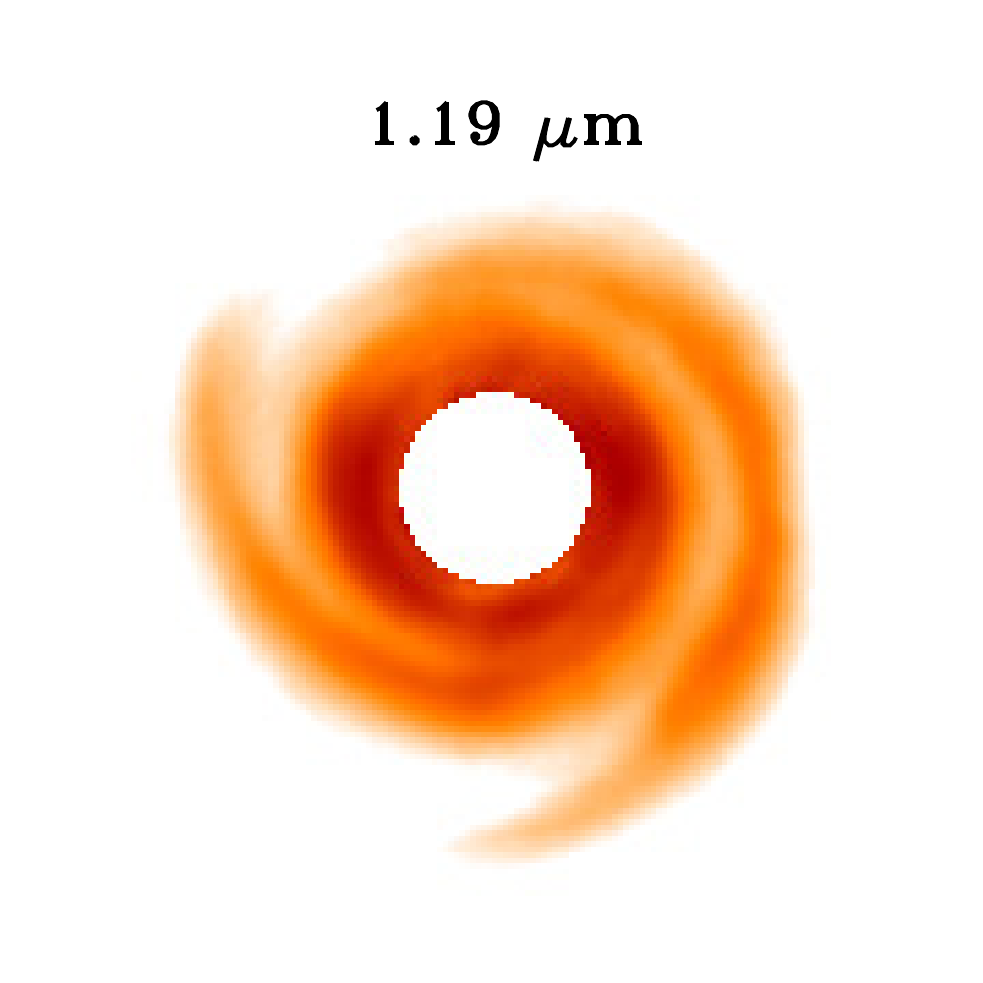}
\includegraphics[trim = 0mm 0mm 0mm 10mm,clip,width=0.175\textwidth]{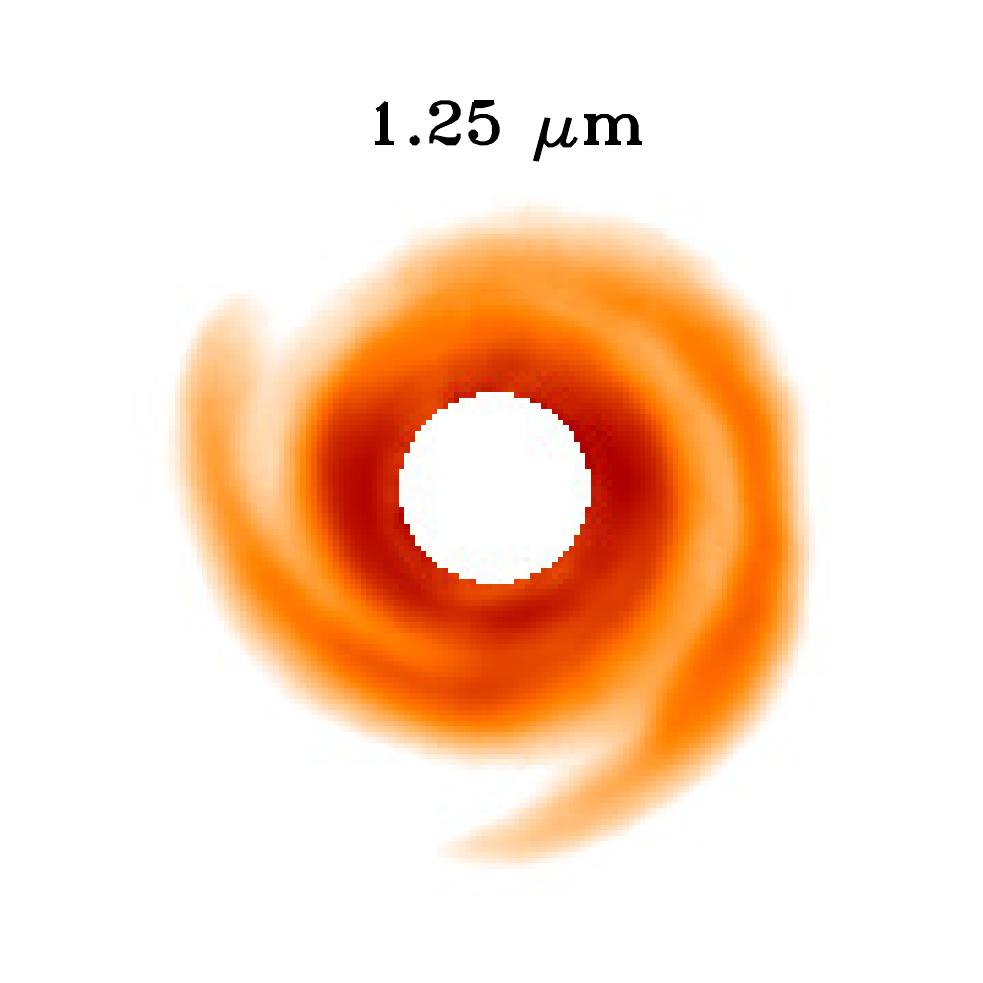}
\includegraphics[trim = 0mm 0mm 0mm 10mm,clip,width=0.175\textwidth]{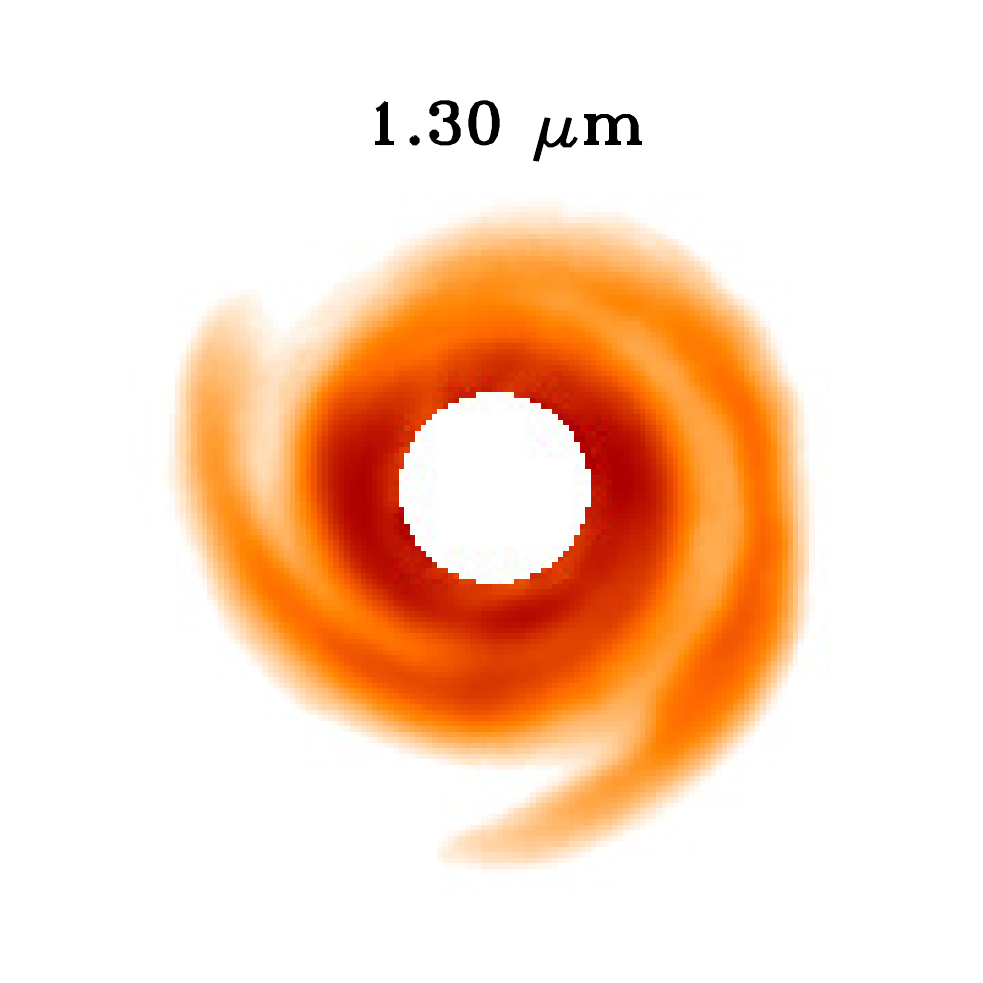}
\includegraphics[trim = 0mm 0mm 0mm 10mm,clip,width=0.175\textwidth]{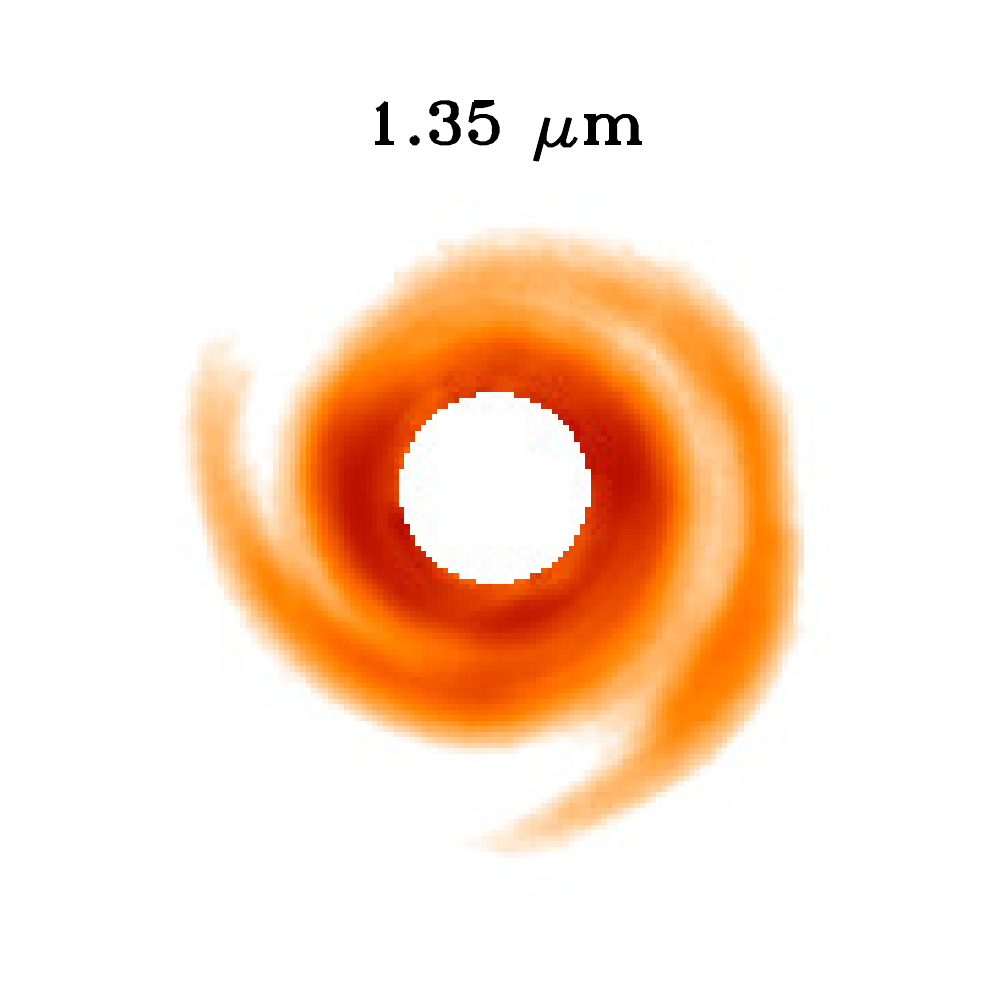}
\includegraphics[trim = 0mm 0mm 0mm 10mm,clip,width=0.175\textwidth]{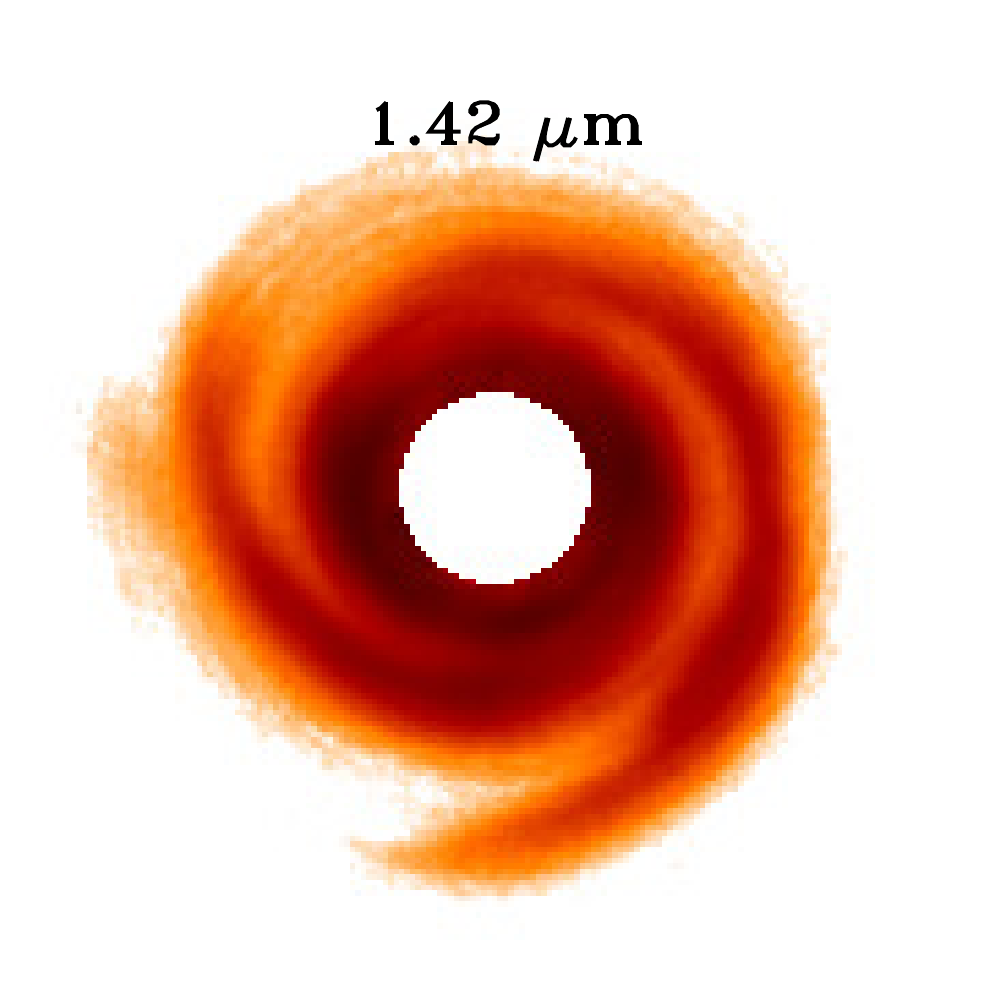}
\includegraphics[trim = 0mm 0mm 0mm 10mm,clip,width=0.175\textwidth]{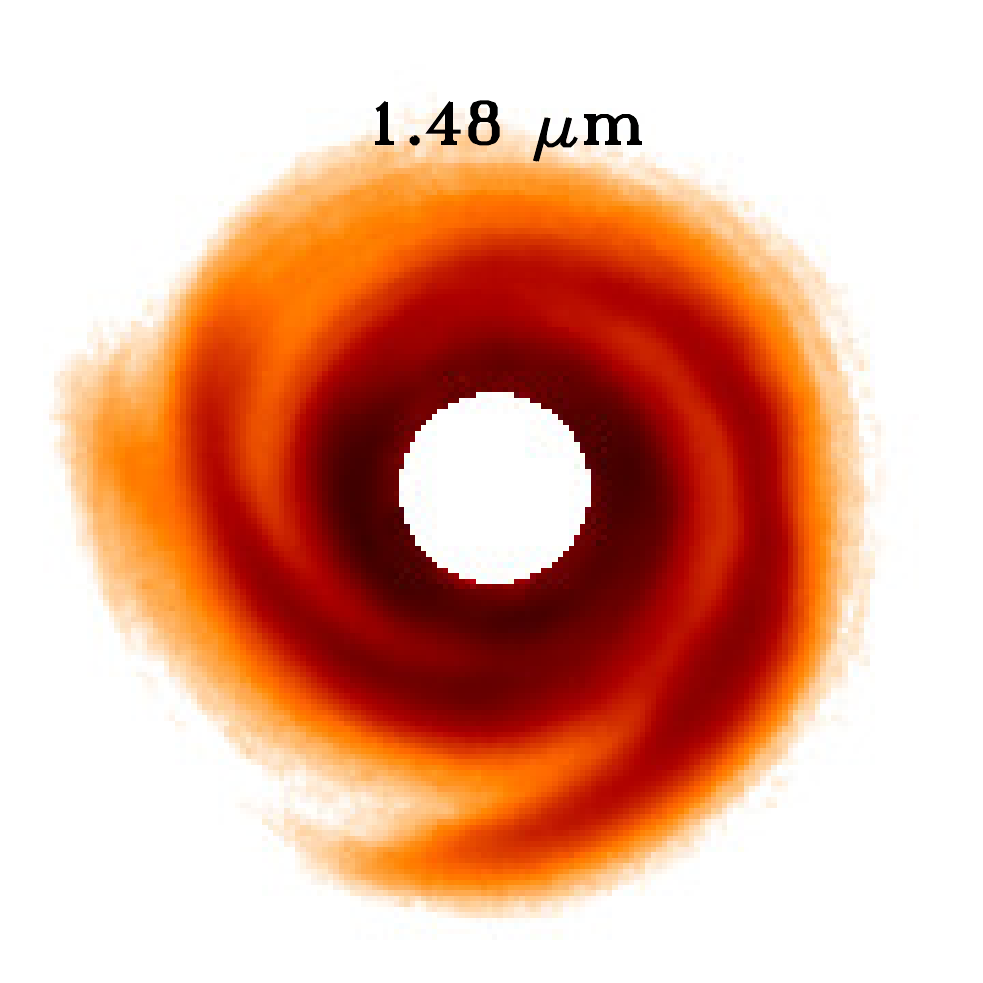}
\includegraphics[trim = 0mm 0mm 0mm 10mm,clip,width=0.175\textwidth]{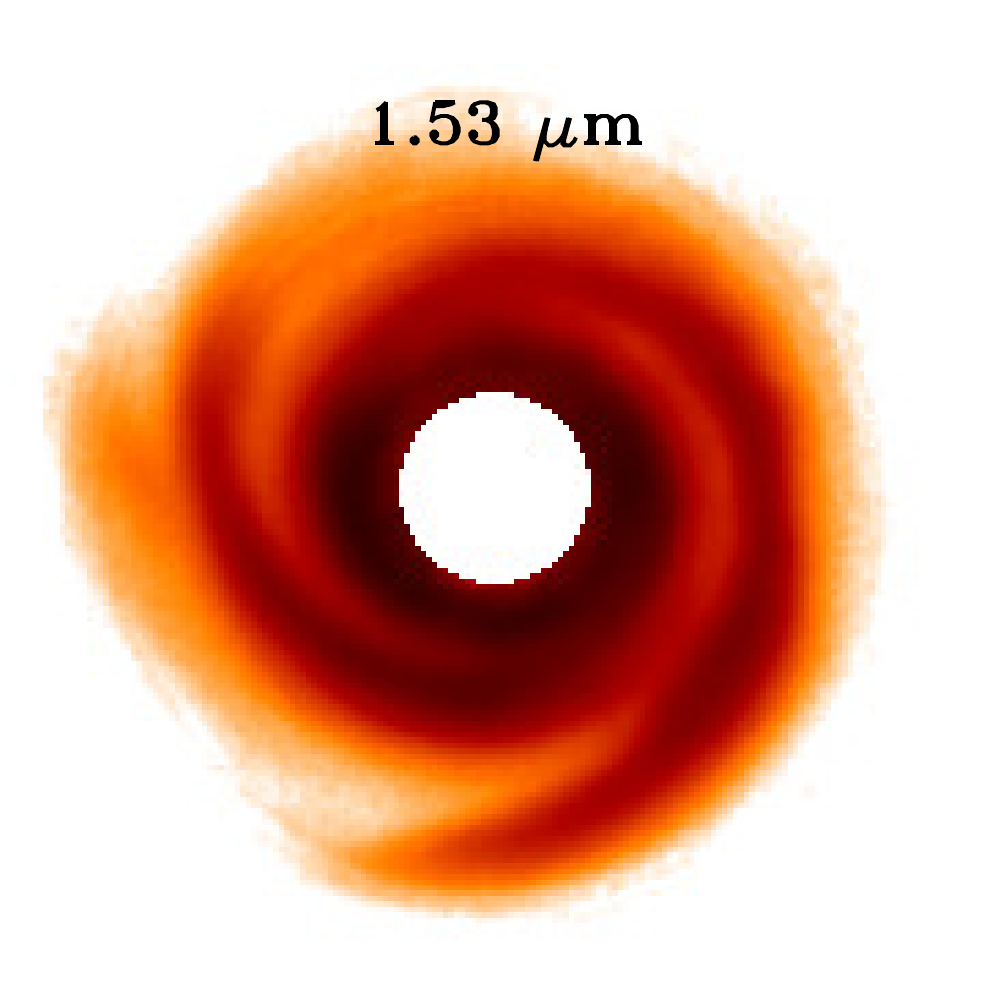}
\includegraphics[trim = 0mm 0mm 0mm 10mm,clip,width=0.175\textwidth]{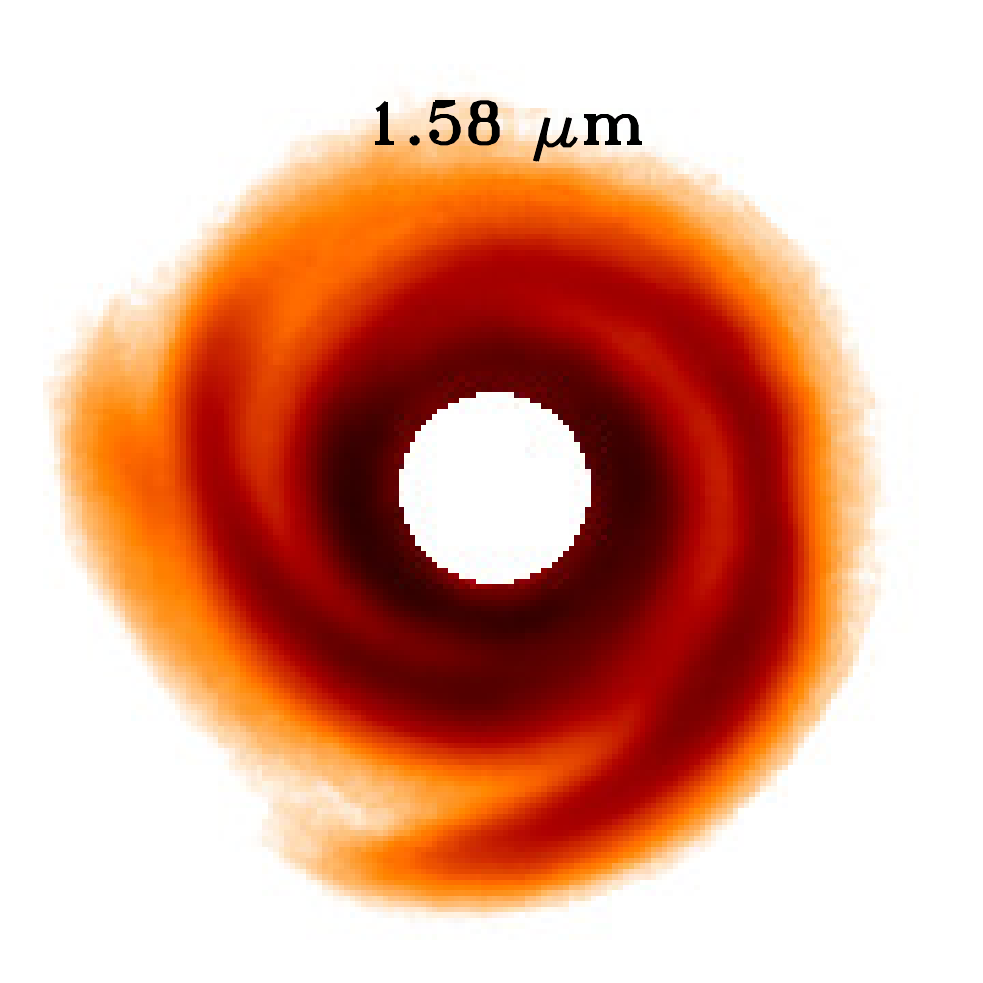}
\includegraphics[trim = 0mm 0mm 0mm 10mm,clip,width=0.175\textwidth]{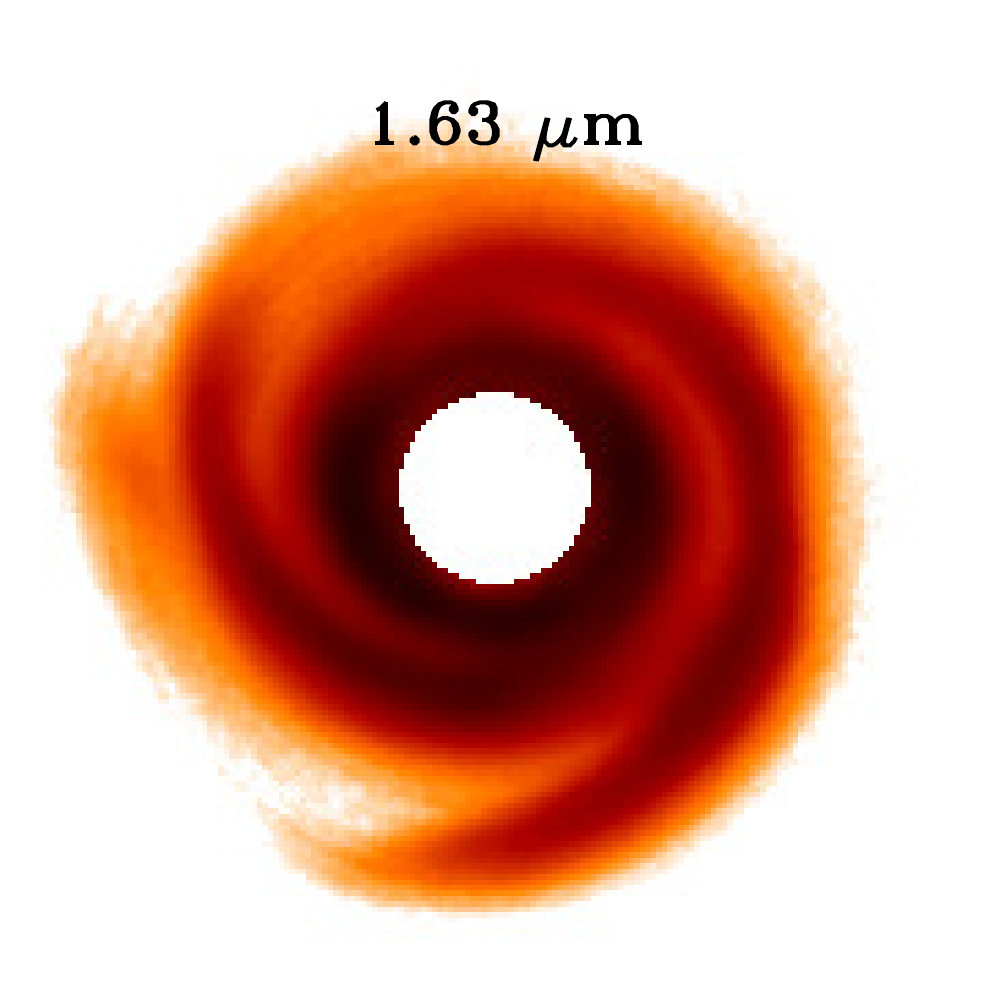}
\includegraphics[trim = 5mm 45mm 5mm 5mm,clip,width=0.75\textwidth]{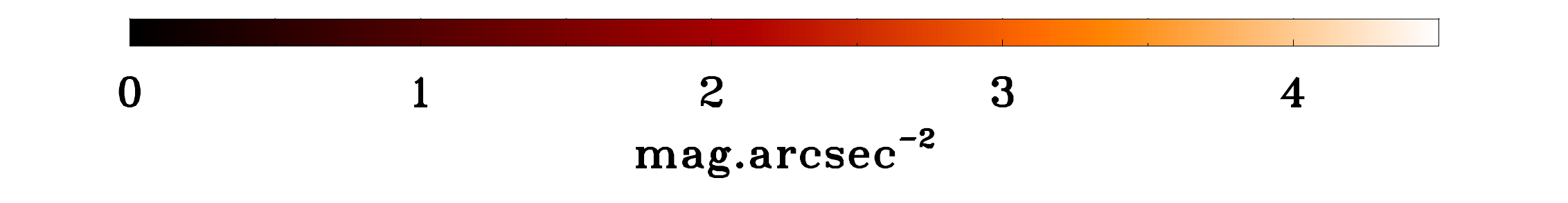}
\caption{{Individual RDI IFS images of SAO~206462 covering the 0.95--1.65~$\muup$m spectral range using a {spectral binning of 3} ($R$\,$\sim$\,9--15). Each image is normalized to the {maximum of the unsaturated non-coronagraphic PSF in each band}. The center of the images is masked out numerically to remove the star.}}
\label{fig:ifsimchannels}
\end{figure*}

Figure~\ref{fig:ifsimchannels} shows the IFS individual images in the $YJH$ bands. We note that the disk does not exhibit morphological spectral variations, except for a global increase in contrast with the wavelength.

\section{Photometric variability of the host star}

\begin{figure*}[t]
\centering
\includegraphics[angle=90,width=0.85\textwidth]{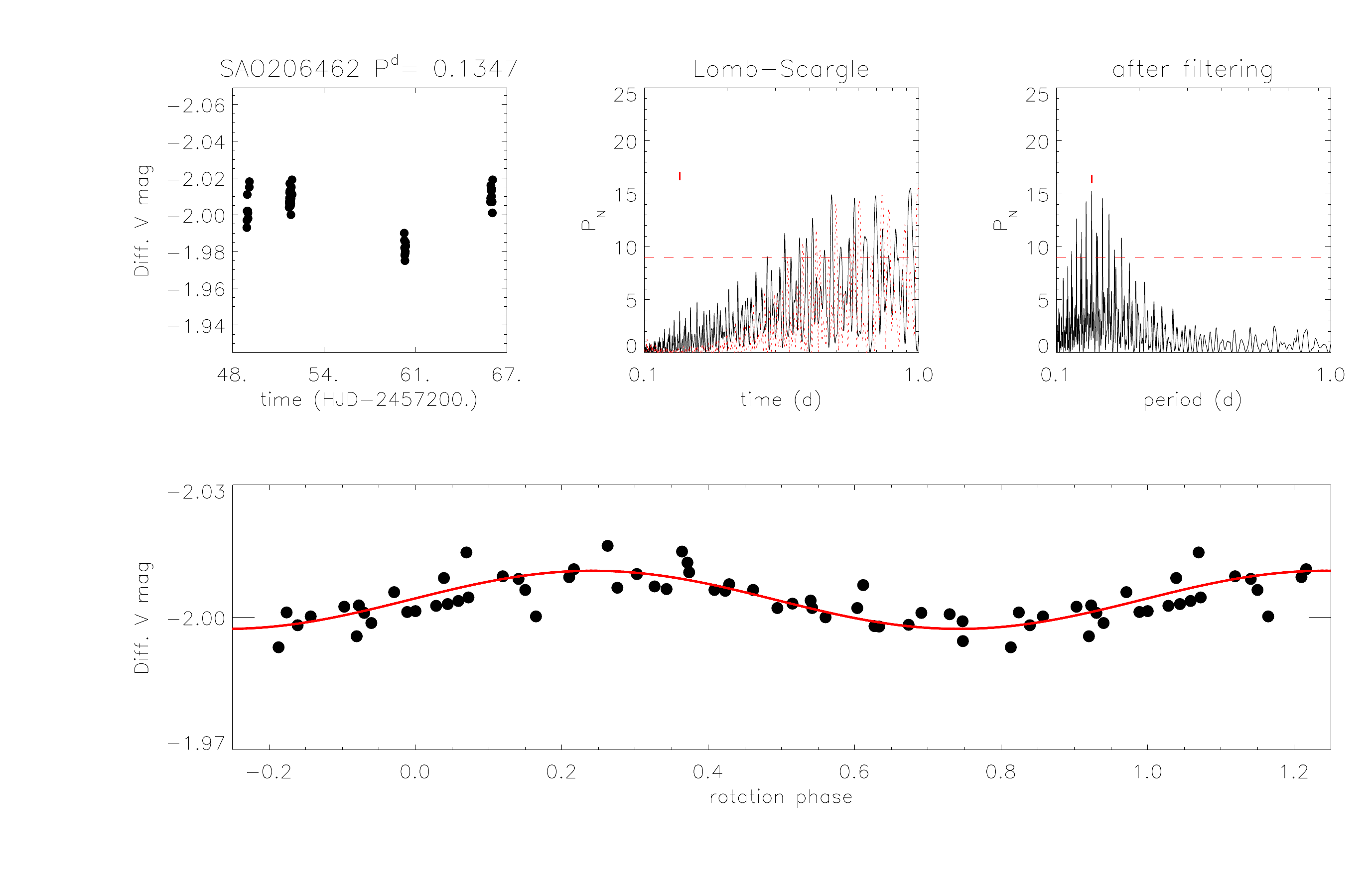}
\caption{Photometric analysis of SAO206462. \textit{Top row from left to right}: $V$-band differential magnitudes versus Heliocentric Julian Day, Lomb-Scargle periodogram, and the periodogram after filtering. For the Lomb-Scargle periodogram, we show the spectral window function (red dotted line), the power level corresponding to FAP = 1\% (horizontal dashed line), and the peak corresponding to the rotation period (red arrow). {\textit{Bottom panel}: Light} curve phased with the rotation period. The solid line represents the sinusoidal fit.}
\label{fig:photometricvar}
\end{figure*}

As supporting observations for the SPHERE observations reported in this paper, we performed a photometric monitoring of the host star in order to analyze its photometric variability. In high-contrast observations, the photometric measurements are relative to the host star, so their accuracy is dependent on the stellar variability. SAO~206462 is known to show high photometric variability \citep[$\sim$23\% at 1.7~$\muup$m,][]{Grady2009}.

The star was observed from August 14 to 31, 2015 for a total of four nights at the York Creek Observatory (41$^{\circ}$06$'$06.4$''$S, 146$^{\circ}$50$'$33$''$E, Georgetown, Tasmania) using a f/10 25\,cm Takahashi Mewlon reflector, equipped with a QSI 683ws-8 camera, and $B$, $V$, and $R$ standard Johnson-Cousins filters. The telescope has a field of view of 24.5$'$\,$\times$18.5$'$. The pixel scale is 0.44$''$/pix. A total of 51 frames were collected in the $V$ filter using an integration time of 20\,s. Aperture photometry was used to extract the magnitudes of SAO~206462 and other stars in the field to be used as comparison stars. All reduction steps were performed using the tasks within IRAF\footnote{IRAF is distributed by the National Optical Astronomy Observatory, which is operated by the Association of the Universities for Research in Astronomy, Inc. (AURA), under cooperative agreement with the National Science Foundation.}. The achieved photometric accuracy was $\sigma_V$\,=\,0.01~mag.

We could identify three stars whose light curves were very stable and therefore suitable as comparison stars. Their differential magnitudes during our observing run were found to be constant within our photometric precision. We performed the Lomb-Scargle periodogram analysis of differential $V$ light curve. In Fig.~\ref{fig:photometricvar} we summarize the results for the $V$-band observations. On the basis of stellar radius $R$\,=\,1.4~$R_{\odot}$ and projected rotational velocity $v$\,sin\,$i$\,=\,82.5~km\,s$^{-1}$ \citep{Mueller2011}, the expected rotation period must be shorter than about 1~day. We searched for photometric periodicities in the 0.1--1.0-day interval. In the periodogram we found a number of significant power peaks that are related to the observation window function. However, after filtering the major power peak, the pre-whitened time series showed a significant periodicity at $P$\,=\,0.137~d that we assume to be the stellar rotation period. In a previous analysis, \citet{Mueller2011} measured a rotation period $P$\,=\,0.160~d using a generalized Lomb-Scargle analysis of a series of FEROS radial velocity measurements covering a time period of 151~days. However, they also found in their periodogram two other peaks of comparable power at $P$\,=\,0.138 and $P$\,=\,0.191~d, the first peak being in good agreement with our estimate.

\end{appendix}

\end{document}